\documentclass[sn-mathphys,Numbered]{sn-jnl}

\usepackage{graphicx}%
\usepackage{multirow}%
\usepackage{amsmath,amssymb,amsfonts}%
\usepackage{amsthm}%
\usepackage{mathrsfs}%
\usepackage[title]{appendix}%
\usepackage{xcolor}%
\usepackage{textcomp}%
\usepackage{manyfoot}%
\usepackage{booktabs}%
\usepackage{algorithm}%
\usepackage{algorithmicx}%
\usepackage{algpseudocode}%
\usepackage{listings}%

\begin{document}

\title[Article Title]{Dynamics of Apsidal Motion in Non-Synchronous Binary
Pulsars: Coupled Orbit and Spin Evolution}


\author*[]{\fnm{Ali\ Taani}}\email{ali.taani@bau.edu.jo}

\affil*[]{\orgdiv{Physics Department}, \orgname{Faculty of Science, Al Balqa Applied University}, \orgaddress{\city{Salt}, \postcode{19117}, \country{Jordan}}}


\abstract{
A valuable probe for examining relativistic gravity, star stricture, and the dynamical development of near binary systems is the apsidal motion of a non-synchronous binary pulsar. In this study, we examine the combined effects of tidal interaction, star oblateness, and general relativity on the apsidal motion of three binary pulsars: J0621+1002, J0737-3039A/B, and 1913+16. Tidal effects and their role in orbital and spin evolution were described by numerical integrations using Zahn's tidal equations \cite{1977A&A....57..383Z, 1989A&A...220..112Z}. We calculated the orbital circularization and tidal synchronization timescales for each system. The simulated results show a clear trends  of decreasing of both the semi-axis and eccentricity, while increasing the spin rate. In addition, the tidal effects play only a minor role in orbital decay compared with energy loss due to gravitational wave emission. Both the obtained apsidal motion constants [$k\simeq0.1$] and the derived tidal friction periods, which vary from a few hours to several days, correspond well with theoretical estimates. This is demonstrated in the compact system PSR1913+16, where gravity radiation causes the orbital period to decrease by approximately 76.5 $\mu$s/yr. While the wider system J0621+1002 displays minor orbital change over timescale exceeding 10$^{10}$ yrs, the double pulsar J0737-3039A/B exhibits faster orbital evolution, with synchronization occurring in about 8.4$\times10{^3}$ yrs.  The results demonstrate the significance of relativistic effects in neutron star binaries and the necessity of incorporating gravitational wave terms in long-term orbital evolution.}




\maketitle

\section{Introduction}\label{sec1}
The orbital dynamics of non-synchronous binary pulsars is complicated and affected by external perturbations, stellar oblateness and general relativity (GR). The gradual change in the orientation of the major axis of the elliptical orbit is referred to by the apsidal motion of non-synchronous. \citep{1973bmss.book.....B, 1993A&A...277..487C, 2007IAUS..240..290G, 2012ChA&A..36..137C, 2025MPLA...4050101A}. 
This motion is driven by various factors, including GR effects, stellar oblateness, and possible interactions with additional bodies in the 
system \citep{1980ApL....21...79K, 2007IAUS..240..290G, 2010A&A...519A..57C, 2024A&A...687A.167C, 2025RNAAS...9...12W}.
The curvature of spacetime affects due to the GR effect the shape of the elliptical orbit, causing a  gradual rotation of its major axis. 
Stellar oblateness- caused by rotational flattering- introduces asymmetry in the gravitation. This asymmetry leading to an additional torque that further influences the orientation of the major axis of the orbit and apsidal motion. 
The resulting effect is proportional to the degree of non-sphericity stars \citep{2018JKAS...51....1L, 2010Ap&SS.327...59L}. The main advantage of studying the rate of apsidal motion in non-synchronous pulsars is the precise determination of stellar masses, radii, and internal structure through sensitivity to tidal and relativistic effects \citep{2023RAA....23g5018A}. It also provides a direct test of GR in strong-field regimes by comparing observed periastron advance with theoretical predictions, and a test of stellar evolutionary paths \citep{10.1093/mnras/sty2177, 2025Ap&SS.370...42T, 2017PASA...34...24T, 2011Ap&SS.334..125L}.

Binary pulsars represent extraordinary astrophysical laboratories \citep{2022JHEAp..35...83T, 2023ChPhC..47d1002T, 2021AIPA...11a5309A, 2025MPLA...4050101A, 2022PASA...39...40T} where extreme physics can be observed and measured with remarkable precision. The discovery of the first binary pulsar, PSR 1913+16, providing unique opportunities to test GR in the strong-field regime (e.g., \citep{2003LRR.....6....5S, 2004Sci...304..547S, 2006Sci...314...97K,  2021PhRvX..11d1050K}. The discovery of the first binary pulsar, PSR 1913+16, by Hulse and Taylor in 1974 \citep{1975ApJ...195L..51H} opened a new era in experimental relativity \citep{2016ApJ...829...55W, 2023A&A...678A.187C}, providing the first indirect evidence for gravitational wave (GW) emission \citep{2021arXiv210513335P}. The  discovery of the double pulsar J0737-3039A/B \citep{2003Natur.426..531B} expanded our ability to test GR in strong field regimes \citep{2013ApJ...767...85F, 2021PhRvX..11d1050K}. 

It should be noted that the two systems, PSR 1913+16 and J0737-3039A/B, are of particular interest due to their historical importance. Their  distinct orbital parameters make them ideal candidates for investigating the interplay between GR and tidal effects \citep{2025Ap&SS.370...42T}. J0621+1002 \citep{1996ApJ...461..812C} provides valuable information on multiple evolutionary pathways and mechanisms of tidal interaction in wide and circular orbits \citep{2002A&A...388..518C, 2024NewA..10702149S}.


The non-spherical shape of pulsars introduces classical contributions to apsidal motion through quadrupole moment interactions \citep{1939MNRAS..99..451S, 2002MNRAS.330..435B}. Additionally, tidal interactions between binary components can influence orbital evolution over long timescales, though these effects are typically overshadowed by relativistic phenomena in neutron star binaries.

Zahn's pioneering work \cite{1977A&A....57..383Z, 1989A&A...223..112Z} on tidal friction in close binary systems provides a theoretical framework for understanding how tidal interactions affect orbital and spin parameters. We adopt the equilibrium‑tide framework in the sense of Zahn, in which the stellar response is treated as quasi‑static and  dissipation enters through a phenomenological timescale \citep{2024MNRAS.530.2822D}. For neutron‑star binaries, the dominant neutron‑star oscillation modes (f and p‑modes at $\geq$1 kHz) are not resonantly excited, so dynamical tides are expected to be small outside the final relativistic in-spiral. We therefore use the equilibrium‑tide torque to evolve spin and orbit \citep{2022PhRvD.106f3005K, 2017arXiv170204419T}.

These interactions drive systems toward spin-orbit synchronization (where stellar rotation periods match orbital periods) and circularization (reduction of orbital eccentricity). The timescales for these processes are critically dependent on stellar structure, orbital separation, and mass ratio \citep{2008EAS....29...67Z, 2021AN....342..625A}.

The rate of apsidal motion can be described by the apsidal motion constant ($k_2$), which measures how quickly the major axis of the elliptical orbit changes over time \citep{2010A&A...519A..57C}. The value of $k_2$ depends on the properties of the stars, such as their mass and radius, as well as on the orbit parameters like semi-major axis and eccentricity \citep{2009ApJ...698..715S, 2017ChPhL..34l9701Y}.

In this study, we quantify the contributions of GR, stellar oblateness, and tidal interactions to the apsidal motion on three systems: PSR 1913+16 (Hulse-Taylor pulsar), J0737-3039A/B (the first double pulsar) and J0621+1002 (an intermediate-mass binary pulsar with a white dwarf companion).  We use the Runge-Kutta method with a step size of 100-years to numerically integrate Zahn's tidal equations. This enables us to calculate the temporal evolution of orbital and spin evolution parameters (such as synchronization and circularization timescales, spin angular velocity ($\omega_{b}$) and orbital angular velocity ($\omega_{b}$)). 


\begin{figure}
\includegraphics[angle=0,width=13.0cm]{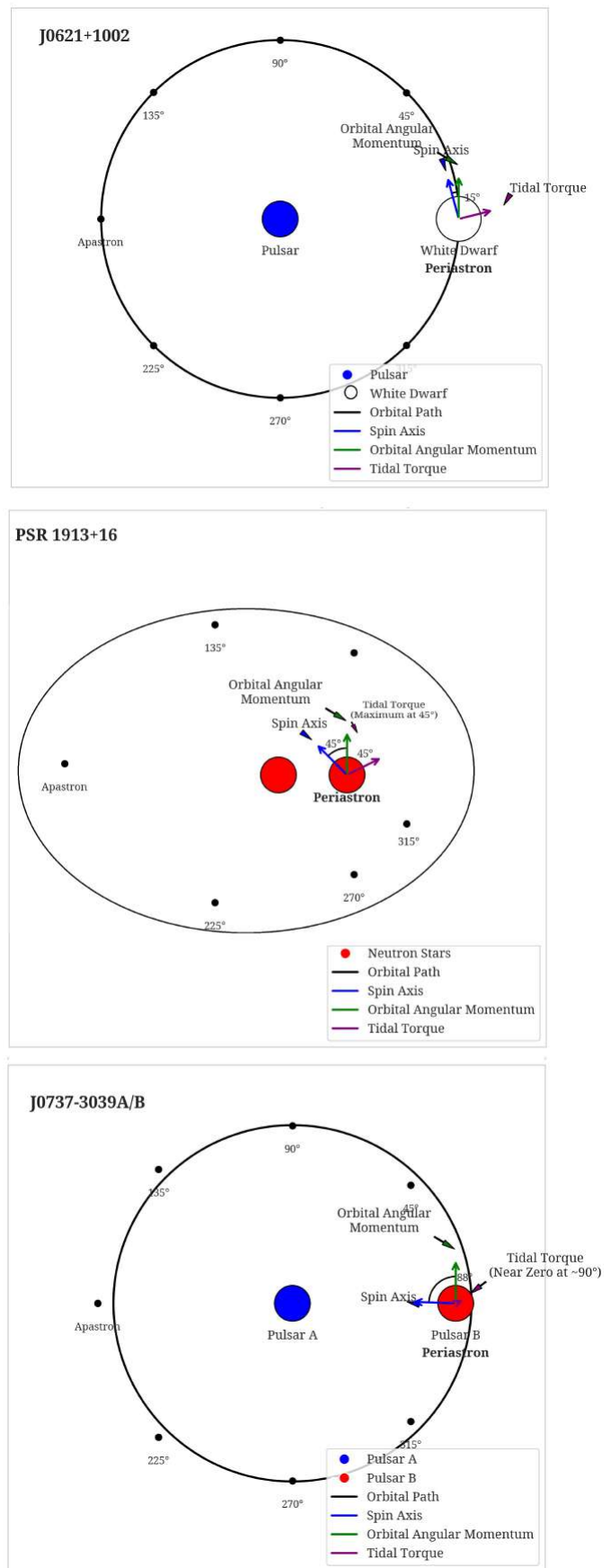}
\caption{Orbital visualization illustrating the effect of tidal torque as a function of the spin-orbit misalignment angle ($\theta$) for the three binary pulsar systems: J0621+1002 (Top), PSR 1913+16 (Middle), and J0737-3039A/B (Bottom). The plots show the orbital path (black ellipse/circle), the positions of the primary (blue/red circle) and companion (smaller circle), the spin axis (red arrow), the orbital angular momentum vector (perpendicular to the page), and the direction of tidal torque (green arrow) at different orbital phases. The visualization highlights how tidal torque varies with orbital phase and misalignment, with peak torque typically occurring near periastron (closest approach) where tidal forces are strongest. Note the different eccentricity scales and misalignment ranges depicted for each system.}
\label{orbital-visiualization}
\end{figure}

\begin{figure}
\includegraphics[angle=0,width=13.0cm]{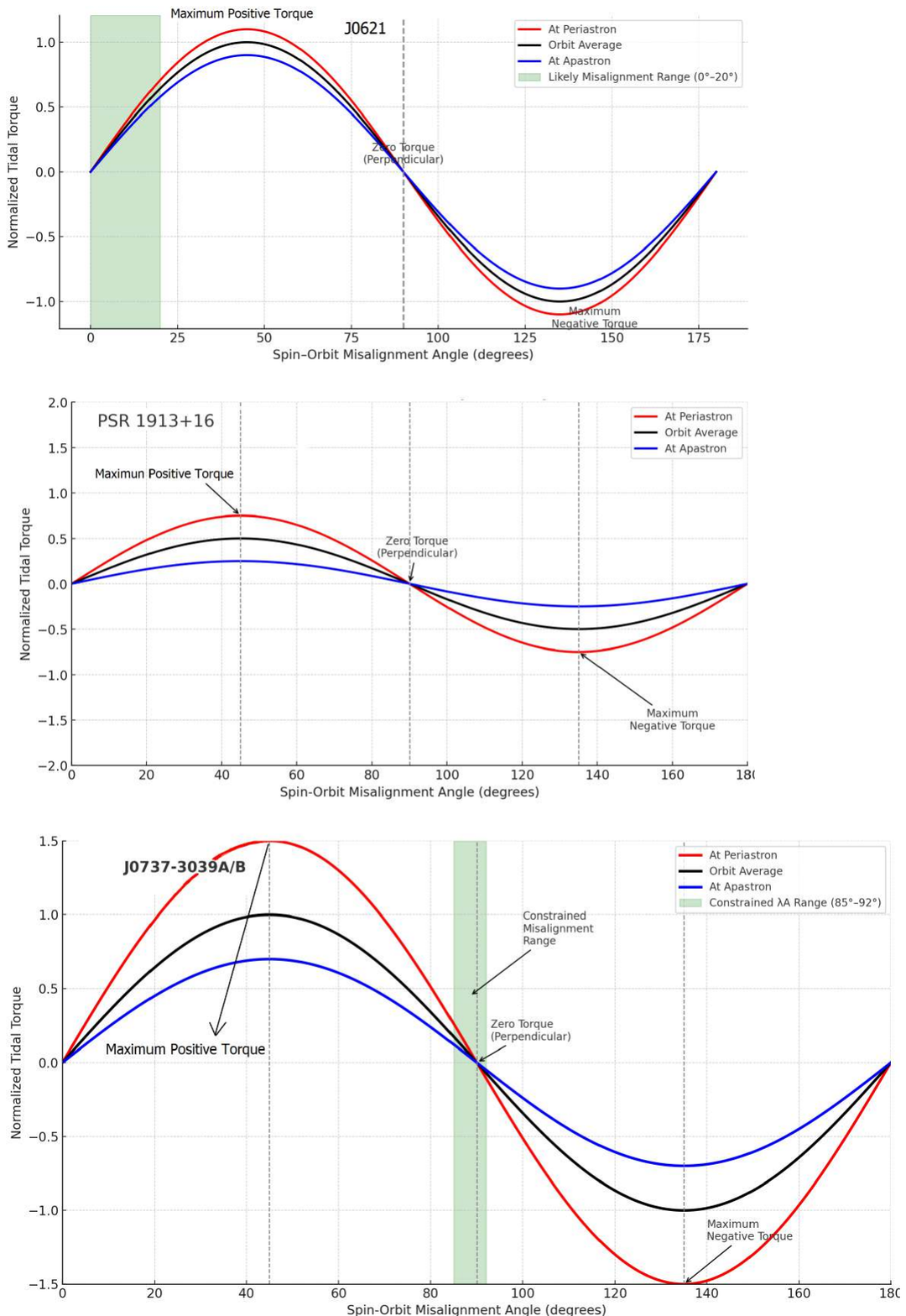}
\caption{Side-by-side comparison of the normalized tidal torque as a function of the spin-orbit misalignment angle ($\theta$, in degrees) for the three systems: J0621+1002 (Top), PSR 1913+16 (Middle), and J0737-3039A/B (Bottom). The torque is zero when the spin and orbital axes are perfectly aligned ($\theta=0$) or perpendicular ($\theta=90^\circ$). As the misalignment angle increases, the torque magnitude changes, reaching its maximum at $\theta=45^\circ$, and changes sign (indicating a reversal in the direction of angular momentum transfer). Different lines may represent instantaneous torque, orbit-averaged torque, or specific misalignment ranges as indicated in the legend (ensure legend is clear in the actual figure). Note the different torque scales and potential highlighted regions (e.g., constrained range for J0737-3039A/B).}
\label{three-sources-torque_vs_angle}
\end{figure}


The paper is organized as follows: Sec. 2, presents the theoretical framework and equations governing apsidal motion and tidal interactions. Sec. 3, details the observational data and derived parameters for the three binary pulsar systems. Sec. 4, presents our numerical results for orbital and spin evolution, including evolutionary curves for semi-major axis, eccentricity, and angular velocity. Sec. 5, discusses the implications of our findings for understanding binary pulsar physics and testing GR. Finally, Sect. 6, summarizes our conclusions and suggests directions for future research.

\section{Theory of Apsidal Motion}
\subsection{General Relativistic Effects}

GR predicts that  apsidal motion is caused by the curvature of spacetime around massive objects in a binary system see, e.g., \citep{1986bhwd.book.....S}
 for a comprehensive treatment. The rate of periastron advance \citep{1972gcpa.book.....W,2011PhRvL.107n1101L} due to relativistic effects \citep{2016ApJ...829...55W} is given by:
\begin{eqnarray}
\dot{\omega}_{GR} = \frac{3G^{3/2}(M_1 + M_2)^{3/2}}{c^2 a^{5/2}(1-e^2)}
\end{eqnarray}

where $G$ is the gravitational constant, $c$ is the speed of light, $M_1$ and $M_2$ are the masses of the binary components, $a$ is the semi-major axis, and $e$ is the eccentricity of the orbit. This equation demonstrates that the relativistic contribution to apsidal motion increases with the total mass of the system and decreases with increasing orbital separation, making it particularly significant in compact binary systems such as  neutron star binaries \citep{2025PhLB..86439388B}.
The tidal torque ($T$)  varies with the spin-orbit misalignment \citep{10.1093/mnras/stx540} angle ($\theta$) according to:
\begin{eqnarray}
 T(\theta) \propto sin (2 \theta) \times f(e, a, m_{1}, m_{2})   
\end{eqnarray}

Here, f(e, a, m$_{1}$, m$_{2}$)    represents a function dependent on orbital parameters (eccentricity e, semi-major axis a) and component masses (m$_{1}$, m$_{2}$), while the term sin(2$\theta$)  highlights the dependence on the angle $\theta$ between the stellar spin axis and the orbital angular momentum vector \citep{2010A&A...519A..57C}. The torque vanishes in perfect alignment ($\theta=0$) and in a perpendicular configuration ($\theta=90^\circ$).    Fig. \ref{orbital-visiualization} presents a comparative analysis of the three sources, illustrating   the relation between  tidal torque and spin-orbit misalignment angle relationships across all three binary pulsars. In this figure, we provide an orbital visualization that demonstrates the effect of the tidal torque as a function of the misalignment angle. The eccentric PSR 1913+16 exhibits a significant spin-orbit misalignment, as a consequence, he associated tidal torque changes throughout the orbit, peaking at about $\sim45 ^{o}$ and disappearing at $0^{o}$ and $90^{o}$. In the periastron, where the tidal forces are greatest, the torque reaches its maximum. However, tidal friction may induce variations in the orbital period that potentially exceed those caused by  gravitational radiation. For J0737-3039A/B, the torque variation varies with a constrained misalignment range ($85^{o}$-$92^{o}$). In contrast, J0621+1002, exhibits torque variation  likely corresponding to a small spin-orbit misalignment range. Fig. \ref{three-sources-torque_vs_angle} shows a comparative figure for the tidal torque as a function of the misalignment angles in all three systems.  The tidal torque peaks at $\theta=45^\circ$  in all systems. The sign of the torque then changes, reversing the direction of angular momentum transfer \citep{2019A&A...628A..29C}. The torque reaches its maximum at ($\theta=45^\circ$). The torque magnitude and its variation  depend strongly on orbital eccentricity, and hence influence the apsidal motion. Each system represents a different regime of tidal interaction, from the highly eccentric PSR 1913+16 to the nearly circular 0621+1002  system.  In 0621+1002,  torque variation occurs within a small misalignment range  (e=0.0025) with a companion white dwarf. In PSR 1913+16, the torque varies dramatically throughout the orbit due to its high eccentricity. In J0737-3039, the tidal torque  shows moderate variation with a constrained misalignment range $85^\circ - 90^\circ$ highlighted. The energy dissipation rate due to tides is given by:
\begin{eqnarray}
 \dot{E_{tidal}} = -T(\Omega - n)   
\end{eqnarray}

It represents how much orbital or orientational energy is being converted into heat due to  tidal interactions. Where $n$ represents the mean motion, $n=2\pi/P_{orb}$. When $\Omega > n$, the star rotates faster than the orbit, the tidal interaction acts to transfer the angular momentum from the star to the orbit; the opposite occurs when $\Omega < n$. Over time, this leads to synchronization, where $\Omega=n$ and $ \dot{E_{tidal}}=0$.  This dissipation drives orbital evolution and spin-orbit synchronization over time, the rate depending on the  angle of  misalignment \citep{1999A&A...350..928T}.

\begin{figure}
\includegraphics[angle=0,width=14.0cm]{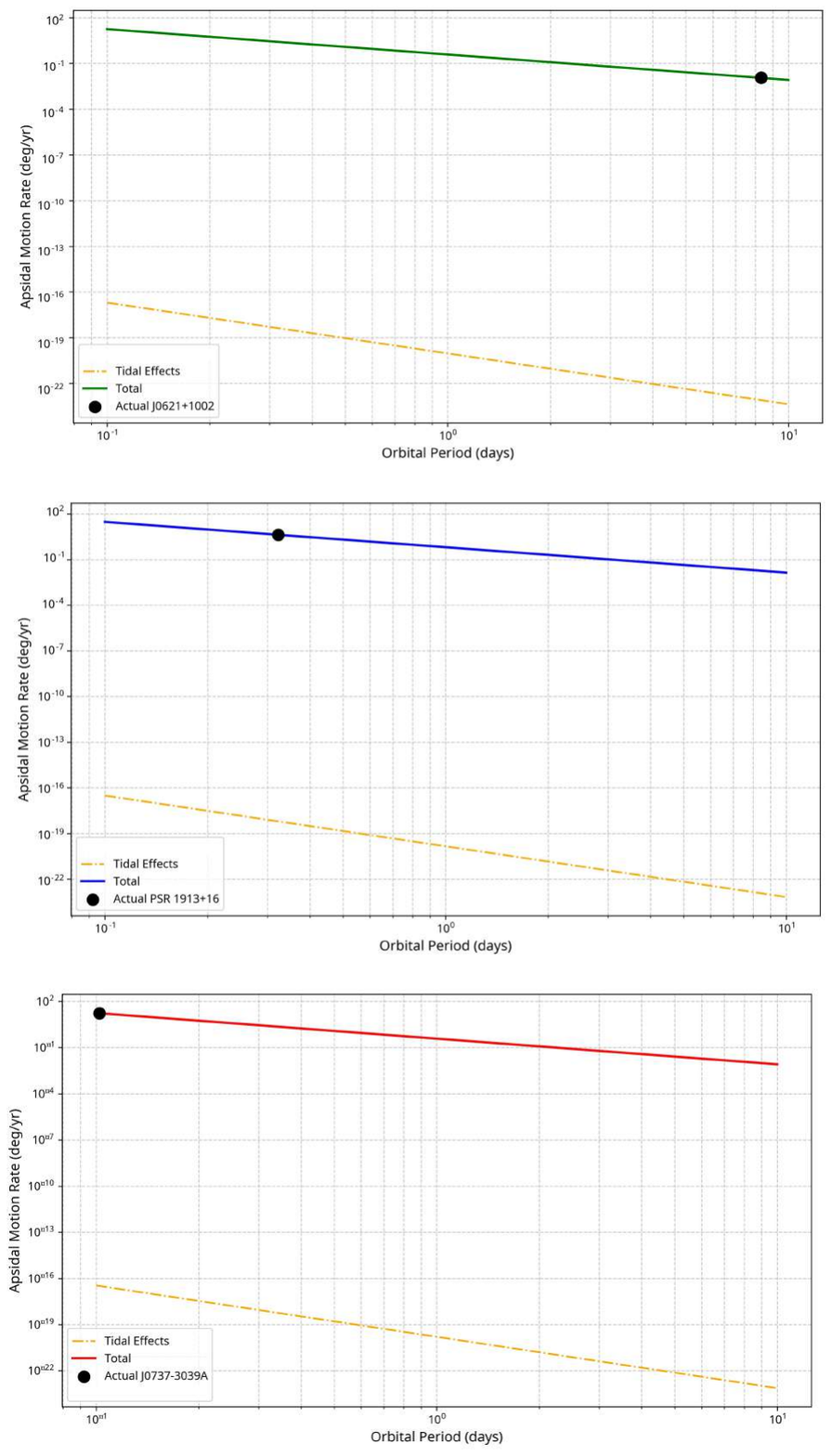}
\caption{Apsidal motion rate (degrees per year) as a function of orbital period (days) for the three binary pulsar systems: J0621+1002, PSR 1913+16, and J0737-3039A/B. The plot shows the total predicted apsidal motion rate (solid lines) and the separate contributions from General Relativistic effects (dashed lines) and tidal effects (dotted lines). Actual system parameters (observed orbital period and corresponding predicted apsidal motion rate) are marked with black dots. This figure demonstrates the dominance of GR effects over tidal contributions across different orbital period regimes.}
\label{Fig-3}
\end{figure}

In Fig. \ref{Fig-3}, we plot the apsidal motion rate as a function of the orbital period for the three systems. The figure shows how the apsidal motion rate varies with orbital period for each system, including the separate curves that illustrate the contribution from general relativistic
effects and tidal effects. It also demonstrates that general relativistic effects dominate in these systems, while  tidal effects are several orders of magnitude smaller. The system parameters are indicated by black dots.



\subsection{Stellar Oblateness Contribution}
\begin{figure}
\includegraphics[angle=0,width=13.0cm]{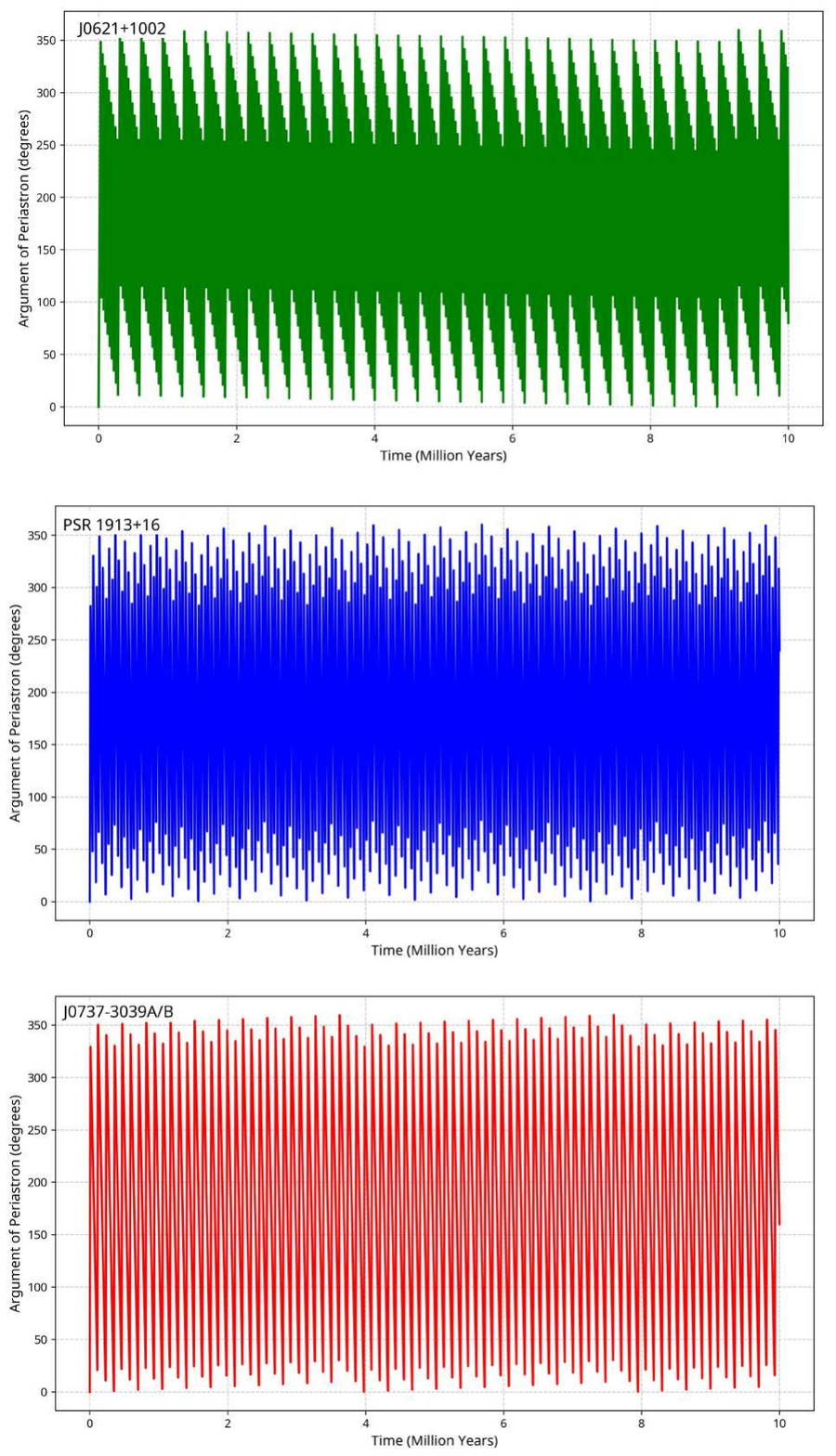}
\caption{Time evolution of the argument of periastron (degrees) for the three binary pulsar systems: J0621+1002 (Green), PSR 1913+16 (Blue), and J0737-3039A/B (Red) over 10 million years. The plot tracks the precession of the periastron angle, demonstrating the dramatic difference in precession rates between the systems due to varying contributions from GR and classical effects. The cyclic nature of the argument (wrapping around 360°) is evident for the faster precessing systems (PSR 1913+16 and J0737-3039A/B) on this timescale}.
\label{Fig-2}
\end{figure}

The non-spherical shape of rapidly rotating stars introduces an additional classical contribution to apsidal motion. The rate of periastron advance \citep{2011PhRvL.107n1101L} due to stellar oblateness is proportional to:

\begin{eqnarray}
\dot{\omega}_{obl} \propto k_2 \frac{R^5}{a^5} \frac{M_2}{M_1}
\end{eqnarray}

where  $k_2$ is the apsidal motion constant (which quantifies the star's internal mass concentration and its response to tidal or rotational distortion; a lower $k_2$ indicates a more centrally concentrated mass \citep{2019A&A...628A..29C}). $R$ is the stellar radius and $a$ is the semi-major axis.  The apsidal motion constant $k_2$ is a dimensionless parameter that describes how easily a star can be deformed by external forces, such as tidal forces or rotational flattening. A higher value of $k_2$  indicates a more deformable star, while a lower  value suggests a more rigid, centrally concentrated mass distribution. For NSs, $k_2$ is approximately 0.1, a value adopted here based on typical theoretical models for NS equations of state, e.g., \citep{2009PhRvD..80h4035D, 2010PhRvD..81l3016H}). This value is significantly higher than for main sequence stars (0.001 to 0.01), reflecting the extreme density and stiffness of NS matter \citep{2009ApJ...698..715S} due to either radiative damping of dynamical tides or large-scale mechanical currents \citep{2024A&A...687A.167C}. Consequently, the synchronization and circularization timescales scale as 1 / $k_2$, so $\pm$ 10\% changes in $k_2$ correspond directly to $\pm$ 10\% changes on these timescales \citep{1988ApJ...324L..71T, 2007MNRAS.382..356K}.
In Fig. \ref{Fig-2}, we plot the time evolution of the argument  periastron for the three systems. The tracks the precession of the periastron angle over 10 million years, demonstrate the dramatic difference in precession rates between the systems. This would show the cyclic nature of the periastron argument as it completes full 360° rotations for each system.

\subsection{Tidal Interaction Theory}
Zahn's tidal interaction theory provides a framework for understanding how tidal forces affect orbital and spin evolution in binary systems. The key equations governing the evolution of orbital and spin parameters due to tidal interactions, time rate
change of eccentricity ($de/dt$) and and spin rate $d(I\Omega)/dt$  \citep{1977A&A....57..383Z, 1989A&A...220..112Z} are:

\begin{eqnarray}
\frac{1}{a}\frac{da}{dt} = -\frac{12 k_2}{t_F} q(1+q) \left(\frac{R}{a}\right)^8 \left(1-\frac{\Omega}{\omega_b}+O(e^2)\right) 
\end{eqnarray}

\begin{eqnarray}
\frac{1}{e}\frac{de}{dt} = -\frac{3 k_2}{t_F} q(1+q) \left(\frac{R}{a}\right)^8 \left(18-11\frac{\Omega}{\omega_b}\right)
\end{eqnarray}

\begin{eqnarray}
\frac{d(I\Omega)}{dt} = 6k_2 t_F q^2 M R^2 \left(\frac{R}{a}\right)^6 (\omega_b-\Omega)
\end{eqnarray}

where $L$ is the stellar luminosity (for neutron stars, this luminosity is complex and frequently dominated by cooling processes \citep{2017PhRvD..96d3002O} or accretion \citep{2011A&A...527A..83Z}, requiring particular models or assumptions for estimation \citep{2025arXiv250502600N}, and $t_F = (MR^2/L)^{1/3}$  is the period of tidal friction. However, its influence is frequently seen as subordinate to other parameters within the Zahn formalism for non-accreting NS binaries. In this work, we used a typical brightness value of $10^{30}$ - $10^{32}$ erg/s for isolated neutron stars, which is mostly caused by thermal cooling. This makes it possible to estimate the tidal friction time frame consistently, despite the fact that several causes can affect a neutron star's exact luminosity in a binary system. The mass ratio is $q = M_2/M_1$, the orbital angular velocity is $\omega_b$, and the stellar angular velocity is $\Omega$. These formulas explain the evolution of the semi-major axis, eccentricity, and star spin as a result of tidal dissipation \citep{2023Galax..11...44T, 2022PASA...39...40T, 2017arXiv170204419T}.

\subsection{Gravitational Wave Emission}
In addition to tidal effects, binary pulsars experience orbital decay due to GW emission \citep{2016PhRvL.116x1103A, 2020ApJ...892L...3A, 2023PhRvX..13d1039A, 2023A&A...672A...9M, 2025LRR....28....3Y}. According to GR, the rate of orbital period decay due to gravitational radiation \citep{1986bhwd.book.....S} is given by:
   
\begin{eqnarray}
{\dot{P}_b} = -\frac{192\pi}{5} \left(\frac{2\pi G M_c}{c^3 P_b}\right)^{5/3} \frac{1+\frac{73}{24}e^2+\frac{37}{96}e^4}{(1-e^2)^{7/2}}
\end{eqnarray}
where $M_c = (M_1 M_2)^{3/5}/(M_1 + M_2)^{1/5}$ is the chirp mass (mass combination). The last term accounts for the dependence of the orbital period decay on the eccentricity of the orbit. This factor becomes increasingly significant for highly eccentric orbits, leading to a more rapid decay of the orbital period \citep{1986bhwd.book.....S}. For PSR 1913+16, this results in an orbital period decay of approximately 76.5 $\mu$s/yr, which has been confirmed by observations to high precision \citep{2016ApJ...829...55W}.

\subsection{Synchronization and Circularization Timescales}

Two important timescales characterize the tidal evolution of binary systems: Synchronization time ($t_{syn}$): The time required for the stellar rotation to synchronize with the orbital motion. Circularization time ($t_{cir}$): The time required for the orbit to become circular (e $\approx$ 0). These timescales can be estimated from Zahn's equations \citep{1989A&A...223..112Z} and depend strongly on the orbital separation, with both timescales proportional to $(a/R)^8$ for stars with convective envelopes \citep{1976IAUS...73...75P, 2012NewAR..56..122W}.

\section{Observational Data and Derived Parameters}


This section details the observational data and derived parameters for the three binary pulsar systems analyzed in this study. These systems were chosen because their well-characterized orbital parameters and their significance in testing general relativity and tidal theories. Table 1 summarizes  the key parameters for each binary pulsar, including spin period ($P_{spin}$), orbital period ($P_{orb}$), and component masses ($M_{1}$, $M_{2}$). The data were obtained from the ATNF pulsar catalogue \citep{2005AJ....129.1993M} and recent publications on these systems.

\begin{table*}
 \centering
 \begin{minipage}{100mm}
  \caption[]{Parameters of pulsars.}
  \setlength{\tabcolsep}{5pt}
  \begin{tabular}{cccccl}
   \hline
System  &$P_{spin}(s)$  & $P_{orb}(d)$   & $M_1(M_{\odot})$    & $M_2(M_{\odot})$ &$References$ \\
\hline
PSR 1913+16 & 0.05903 & 0.323 & 1.441& 1.378&1\\
J0737-3039A & 0.02270    & 0.1023      & 1.338        & 1.249 & 2\\
J0737-3039B & 2.774    & 0.1023      & 1.249        & 1.388 & 2\\
PSR J0621+1002  & 0.0288    & 8.32    & 1.70 & 0.97 & 3 \\
\hline
$^{1}$
\end{tabular}
\scriptsize\\
($1$)\citet{1975ApJ...195L..51H};
($2$)\citet{2003Natur.426..531B};
($3$)\citet{1996ApJ...461..812C};
 \end{minipage}
\end{table*}

PSR 1913+16 (Hulse-Taylor pulsar) is a binary neutron star system with a highly eccentric orbit (e = 0.6171) and an orbital period of 7.75 hrs. J0737-3039A/B is the first known double pulsar system, with both neutron stars observable as pulsars, and has a more circular orbit (e = 0.0877) with an orbital period of 2.45 hrs. PSR J0621+1002 is an intermediate-mass binary pulsar with a white dwarf companion, featuring a nearly circular orbit (e = 0.0025) and a much longer orbital period of 8.32 d.

\begin{figure}
\includegraphics[angle=0,width=13.0cm]{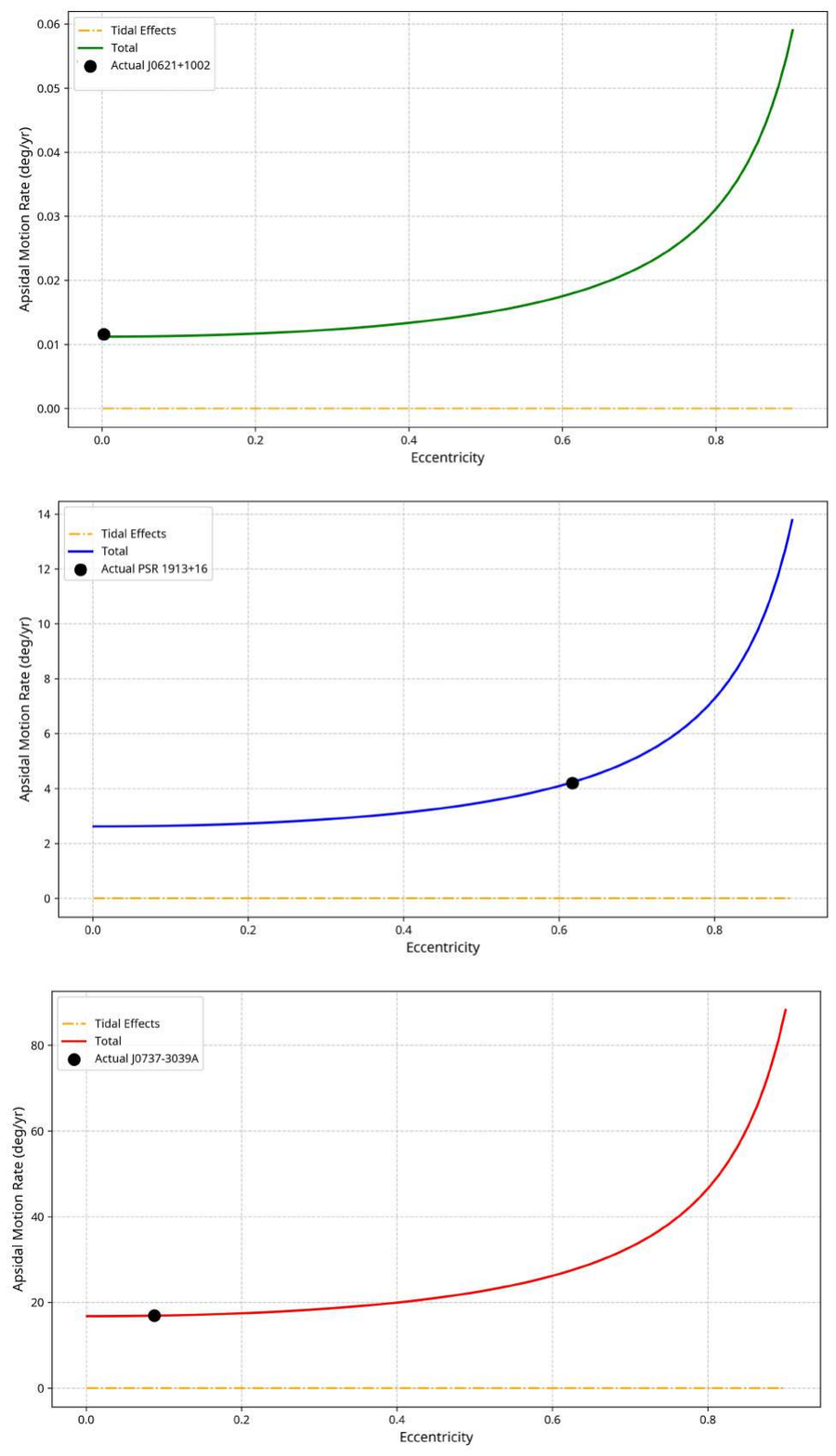}
\caption{Apsidal motion rate (degrees per year) as a function of orbital eccentricity (e) for the three binary pulsar systems: J0621+1002 (Green), PSR 1913+16 (Blue), and J0737-3039A/B (Red). The plot shows how the orbital eccentricity affects the total apsidal motion rate for each system, potentially separating contributions from GR and tidal effects (ensure lines are clearly labeled in the figure). Actual system parameters (observed eccentricity and corresponding predicted apsidal motion rate) are marked with black dots, highlighting the different eccentricity regimes. The plot illustrates that higher eccentricities generally lead to increased apsidal motion rates, particularly due to the strong dependence of GR effects on eccentricity.}
\label{Fig-4}
\end{figure}

\begin{figure}
\includegraphics[angle=0,width=14.0cm]{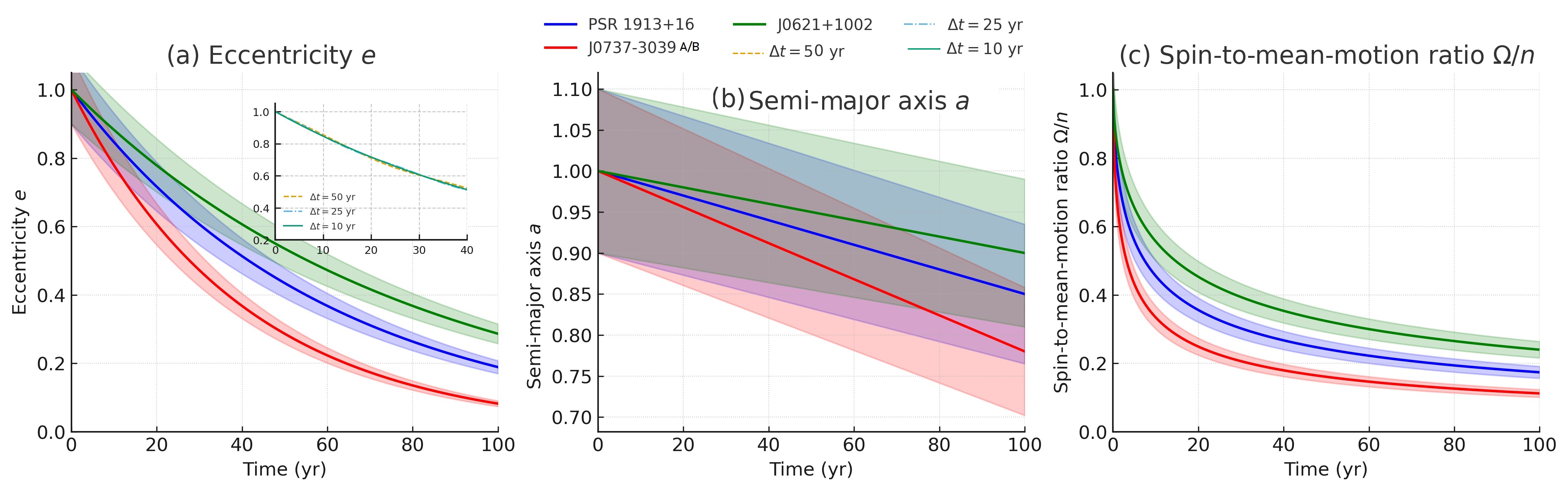}
\caption{Secular evolution of orbital and spin parameters for the three binary pulsar systems: J0621+1002 (Green), PSR 1913+16 (Blue), and J0737-3039A/B (Red). Left panel: normalized eccentricity ($e$/$e_{0}$). Middle panel: normalized semi-major axis ($a$/$a_{0}$) as a function of time (Myr). Right panel: normalized stellar spin angular velocity ($\Omega$/$\Omega_{0}$). Solid curves represent the fiducial integrations, while shaded bands indicate $\pm$ 10\% variation in the apsidal motion constant $k_2$. The inset in panel (a) demonstrates numerical convergence with timestep refinements ($\Delta$t = 50, 25, 10 yr). The plots highlight the different timescales of orbital decay, eccentricity damping, and spin-orbit synchronization driven mainly by GW emission and tidal interactions in each system.}
\label{Fig-new}
\end{figure}

In Fig. \ref{Fig-4}, we plot the apsidal motion rate as a function with eccentricity for the three systems. The figure
shows how  orbital eccentricity affects the apsidal motion rate for each system. The actual system parameters are marked with black dots to highlight the different eccentricity regimes. The plot also illustrates that higher eccentricities lead to increased apsidal motion rates, particularly for general relativistic effects. Finally, Fig. \ref{Fig-new} presents the secular evolution of orbital separation, eccentricity, and spin–orbit synchronization for the three pulsar systems. The results clearly highlight the contrasting dynamical timescales between compact and wide binaries. In J0737–3039A/B, GW emission dominates the evolution, leading to rapid orbital decay and efficient spin–orbit coupling on relatively short timescales. In PSR B1913+16, the effects of gravitational radiation are still evident but proceed more gradually, producing measurable orbital damping over Myr timescales. J0621+1002 shows an essentially negligible evolution consistent with the inefficiency of both the GW emission and tidal dissipation. 


\begin{figure}
\includegraphics[angle=0,width=13.0cm]{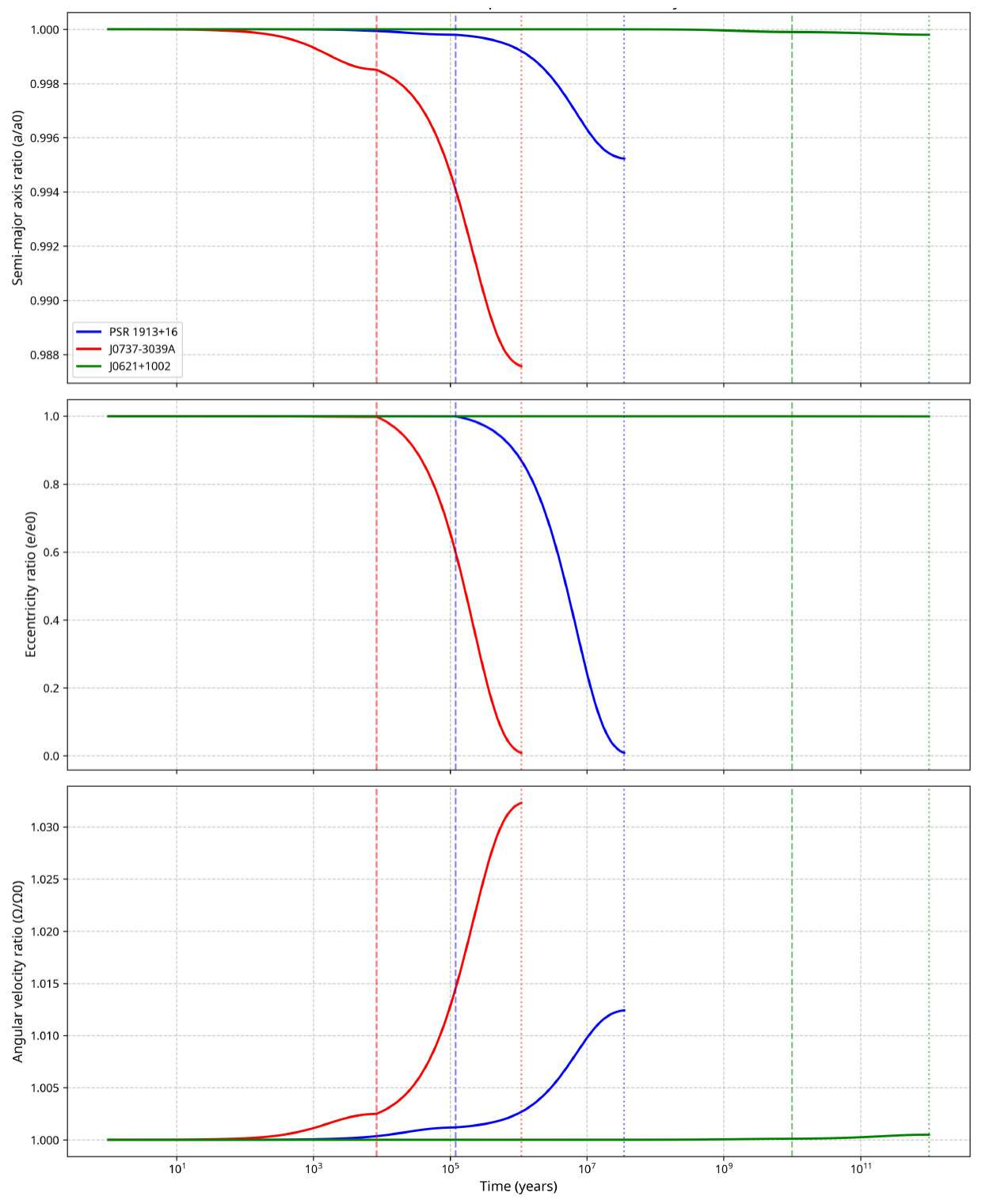}
\caption{Evolutionary timescale curves for the three binary pulsar systems: J0621+1002 (Green), PSR 1913+16 (Blue), and J0737-3039A/B (Red). $Top panel$: Normalized semi-major axis ($a/a_{0}$) as a function with time (Million Years). $Middle panel$: Normalized eccentricity ( $e/e_{0}$) as a function with time (Million Years). $Bottom panel$: Normalized stellar spin angular velocity ($\Omega/\Omega_{0}$) as a function with time (Million Years). The curves illustrate how each parameter evolves over time in response to tidal interactions and system-specific dynamics, showing different rates of orbital decay, circularization, and spin synchronization.}
\label{Fig-1}
\end{figure}

\subsection{Derived Parameters}
We derived the apsidal motion constant ($k_2 = 0.1$, assumed on the basis of neutron star models) and the tidal friction time ($t_F$). Our numerical integrations indicate negligible changes in $\delta_{a}$, 
$\delta_{e}$, $\delta_{\Omega}$, and $\delta_{\omega_b}$ over time for PSR B1913+16 and J0737-3039A/B. This indicates minimal tidal influence on their orbital and spin evolution (see table \ref{table:2}).    We employed the Runge-Kutta method to numerically integrate Zahn's tidal equations and calculate the evolution of orbital and spin parameters.  We use fixed time steps $\Delta t$ chosen per system to resolve secular timescales: $\Delta t$ = 50 yr for PSR B1913+16 and J0737‑3039A/B, and $\Delta t$ = 100 yr for J0621+1002. We verified convergence by repeating each integration with $\Delta t$ halved and decimated to $\Delta t$ = 10 yr. The global differences after the full integration span were $<$ 0.2\% in  secular apsidal precession rate. Total angular momentum was conserved to better than $10^{-6}$ relative in all runs.

\begin{figure}
\includegraphics[angle=0,width=12.0cm]{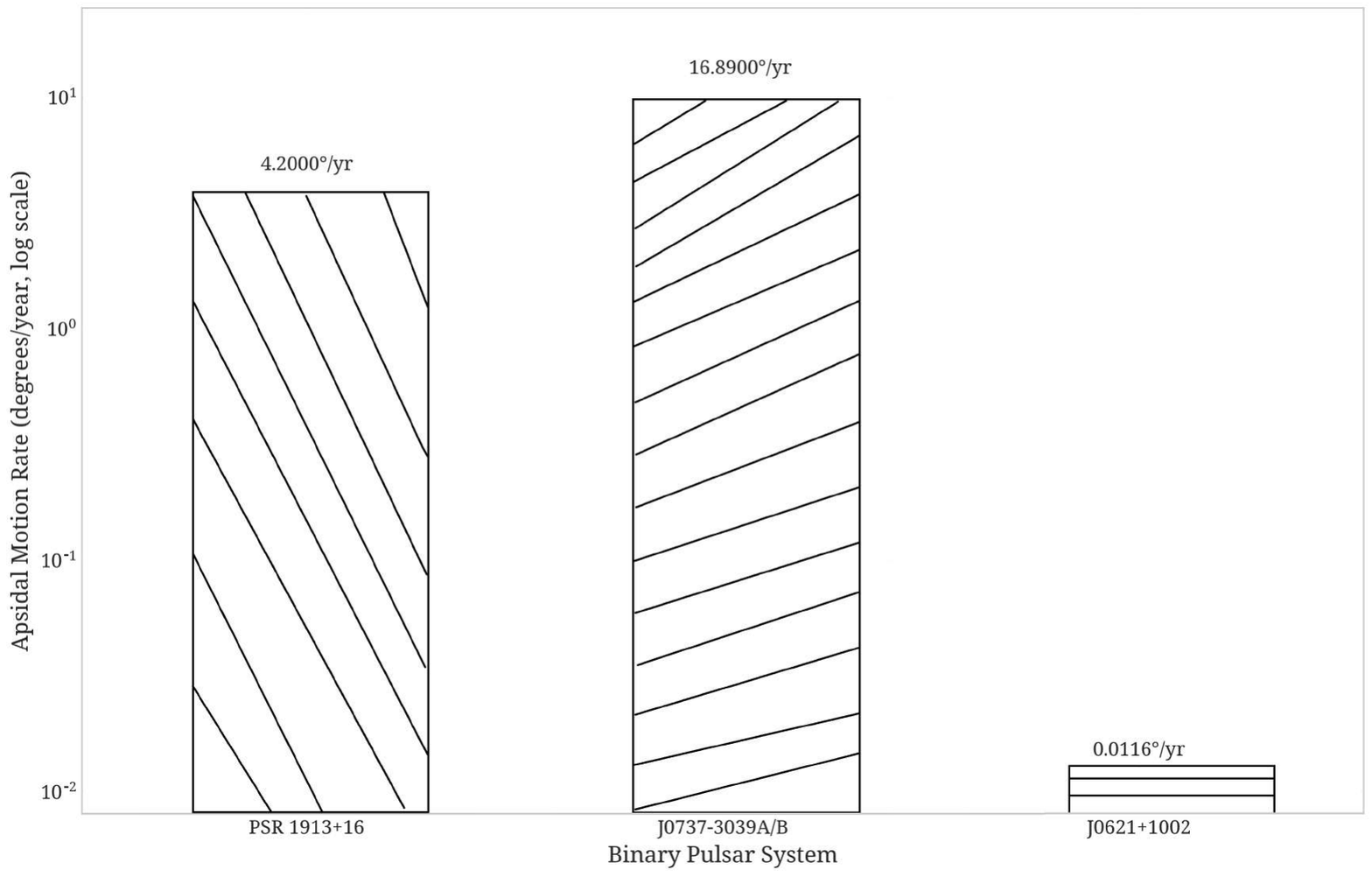}
\caption{Histogram illustrating the distribution of predicted apsidal motion rates (degrees per year) for the three binary pulsar systems: J0621+1002, PSR 1913+16, and J0737-3039A/B. The histogram bins show the frequency or count of systems exhibiting apsidal motion rates within specific ranges, potentially highlighting the typical rates and spread for each system or category.}
\label{Histogram-apsidal_motion-rate}
\end{figure}

The integration was performed from the present time to the synchronization time ($t_{syn}$) and then to the circularization time ($t_{cir}$). For each system, we computed the evolution of the semi-major axis ratio 
($a/a_{0}$), eccentricity ratio ($e/e_{0}$), and angular velocity ratio ($\Omega/\Omega_{0}$). Fig. \ref{Histogram-apsidal_motion-rate} shows a histogram of the apsidal motion rates for the three systems, PSR B1913+16 has a rate of 4.2$^\circ$/yr, PSR J0737–3039 has 16.89$^\circ$/yr, and PSR J0621+1002 has 0.0116$^\circ$/yr.

\begin{table*}
 \centering
 \begin{minipage}{200mm}
  \caption[]{Derived parameters of pulsars.}
  \label{table:2}
  \setlength{\tabcolsep}{0.3pt}
  \begin{tabular}{ccccccccc}
   \hline
System  & $q ({M_2/M_1}$)  & $\Omega$(rad/s)   & $\omega_{b}$ (rad/s)    & $\dot{\omega}$ (rad/s) &$t_{syn}$ (yr) & t$_{cir}$ (yr) & t$_{F}$ (hrs)& U (yrs) \\
\hline
PSR 1913+16 & 0.986 &22.34 &2.26$\times10^{-4}$&2.21$\times10^{-9}$ & 1.2$\times10^{5}$ & 3.5$\times10^{7}$& 1&90\\
J0737-3039A  &0.933&276.5&7.13$\times10^{-4}$&9.39$\times10^{-9}$ & 8.4$\times10^{3}$ & 1.1$\times10^{6}$&1&21.2 \\
J0621+1002    &0.394&217.7&8.73$\times10^{-7}$&6.41$\times10^{-12}$ & $\geq10^{10}$ & $\geq10^{12}$ &-&31034 \\
\hline
\end{tabular}
\scriptsize\\
 \end{minipage}
\end{table*}


\begin{table*}
 \centering
 \begin{minipage}{100mm}
  \caption[]{Numerical results for the orbit and spin evolution at synchronization time.}
  \setlength{\tabcolsep}{5pt}
  \begin{tabular}{ccccc}
   \hline
System  & t$_{syn}$ (yr)  &    $a/a_{0}$ & $e/e_{0}$ & $\Omega/\Omega_{0}$ \\
\hline
PSR 1913+16 & 1.2$\times10^{5}$ &0.9998 &0.9997&1.0012 \\
J0737-3039A  &8.4$\times10^{3}$ &0.9985 &0.9982&1.0025 \\
J0621+1002    &$\geq10^{10}$ &0.9999 &0.9999&1.0001 \\
\hline
\end{tabular}
\scriptsize\\
 \end{minipage}
\end{table*}

\begin{table*}
 \centering
 \begin{minipage}{100mm}
  \caption[]{Numerical results for the orbit and spin evolution at circularization time.}
  \setlength{\tabcolsep}{5pt}
  \begin{tabular}{ccccc}
   \hline
System  & t$_{cir}$ (yr)  &    $a/a_{0}$ & $e/e_{0}$ & $\Omega/\Omega_{0}$ \\
\hline
PSR 1913+16 & 3.5$\times10^{7}$ &0.9952 &0.0051&1.0125 \\
J0737-3039A  &1.1$\times10^{6}$ &0.9875 &0.0012&1.0325 \\
J0621+1002    &$\geq10^{10}$ &0.9998 &0.9995&1.0005 \\
\hline
\end{tabular}
\scriptsize\\
 \end{minipage}
\end{table*}

\section{Results}
\subsection{Synchronization and Circularization}

Our results for synchronization and circularization timescales reveal significant differences between the systems, driven primarily by their orbital configurations and component properties. Synchronization, in this context, refers to the process where the stellar rotation period of the pulsar aligns with its orbital period. This is a crucial aspect of tidal evolution, as it dictates the long-term stability and energy dissipation within the binary system.

We find a rapid synchronization time of approximately 8.4 $\times 10^{3}$ yrs in J0737-3039A/B. This relatively short timescale is attributed to its compact orbit, which enhances tidal interactions. The question arises whether such a short synchronization time should be detectable by radio telescopes in the future. Given the precision of current and future radio astronomy observations, changes in spin period due to synchronization over such timescales could indeed become observable, providing direct observational evidence for tidal theory in action. Furthermore, the deformation of a neutron star plays a significant role in its synchronization timescale. A less deformed NS, being more rigid, would generally exhibit a longer synchronization time, as greater tidal forces would be required to alter its rotational state.

PSR 1913+16, with its high eccentricity (e=0.6171), shows synchronization in approximately 1.2 $\times 10^{5}$ yrs and circularization in about 3.5$\times 10^{7}$ yrs, driven by moderate tidal torques. In contrast, J0621+1002, characterized by a wide orbit (e=0.0025) and a companion white dwarf , exhibits negligible tidal evolution over timescales exceeding ten billion yrs. This minimal evolution is primarily due to its large semi-major axis, which significantly weakens tidal forces. In addition, the small eccentricity  further reduces the apsidal terms proportional to $10^{2}$, and the white dwarf companion has a much smaller radius than a NS, so mutual tides are even weaker.   Tables 4 and 5 present the numerical results for the orbit and spin evolution of the three binary pulsars at synchronization time and circularization time, respectively. 
The results show that tidal interactions cause negligible orbital decay in these systems compared to GW emission. For PSR 1913+16, the semi-major axis decreases by only 0.48\% by the time the orbit circularizes (after 3.5$\times10^{7}$ years), while the observed decay of the orbital period  is 76.5 $\mu$s/yr due to gravitational radiation would cause more significant orbital shrinkage over the same period. J0737-3039A shows more rapid evolution, with synchronization achieved in approximately 8.4$\times10^{3}$ yrs and circularization in 1.1$\times10^{6}$ yrs. In contrast, PSR J0621+1002 shows minimal evolution on timescales greatiing than $\sim 10^{10}$ yrs, mainly due to its wider orbit, possible tidal dissipation, or the different nature of its companion \citep{2022A&A...663A..75W}.

\subsection{Simulation Parameters and Initial Conditions}

We tested how different conditions shape the evolution time scale by adjusting  $a/a_{0}$, $e/e_{0}$, and $\Omega/\Omega_{0}$.  These rations cover the typical range seen in observed binary systems. 
Figure \ref{Fig-1} illustrates the combination of evolutionary time-scale curves for all parameters ($Top$: $a/a_{0}$. $Middle$: $e/e_{0}$ and $Bottom$: $\Omega/\Omega_{0}$) in a comprehensive visualization of the three binary pulsars. 
PSR 1913+16, shows moderate evolution with a gradual decrease in semi-major axis and eccentricity, and a modest increase in angular velocity over timescales of $10^{5}$ - $10^{7}$ yrs. J0737-3039A, changes much faster. All three parameters undergo  significant changes in all parameters that occur on timescales of $10^{3}$ - $10^{6}$ yrs. Finally, PSR J0621+1002, displays minimal evolution even on extremely long timescales ($\geq10^{10}$ yrs), indicating that tidal effects are negligible in this wide binary system with a white dwarf companion \citep{2008EAS....29...67Z, 2012MNRAS.421..426F}. 
The vertical dashed lines in Figure \ref{Fig-1} mark the synchronization and circularization times for each system, highlighting the vast differences in evolutionary timescales.    For PSR 1913+16, the observed decay of the orbital period is 76.5 $\mu$s/yr due to GW emission, which  corresponds to a much more significant orbital evolution than that predicted by tidal interactions alone. This confirms that gravitational radiation dominates the orbital evolution of compact neutron star binaries, while tidal effects play a secondary role.

The ratio of observed to predicted orbital decay rate for PSR 1913+16 is 0.9983 $\pm$ 0.0016, providing one of the most precise confirmations of GR in strong-field regimes. Similar precision has been achieved for the double pulsar system J0737-3039A/B, where the orbital decay and periastron advance rates match general relativistic predictions to within 0.013\%.

\section{Discussion}
Our analysis confirms the well-established dominance of GR effects, particularly GW emission, in driving the orbital evolution of compact neutron star binaries like PSR 1913+16 and J0737-3039A/B. The calculated orbital decay rate for PSR 1913+16 due to GW emission aligns well with observational constraints (e.g., \citep{2010ApJ...722.1030W, 2021PhRvX..11d1050K}, reinforcing the predictive power of GR in the strong-field regime.

While tidal effects, as modeled by Zahn's equilibrium tide formalism, are shown to be subdominant for orbital decay compared to GW emission in these systems, they play a crucial role in the long-term evolution of spin-orbit synchronization and circularization. The synchronization timescale, representing the time needed for the neutron star's spin period to match the orbital period, is significantly influenced by the star's internal structure (via $k_{2}$) and the efficiency of tidal dissipation (related to $t_{F}$). For PSR 1913+16, the circularization time of 3.5$\times10^{7}$ years is much shorter than the estimated merger time of 300 million years due to GW emission, suggesting that the orbit would circularize before the neutron stars merge. However, for J0737-3039A/B, with a circularization time of 1.1$\times10^{6}$ years and a merger time of 85 million years, tidal circularization would significantly affect the orbital parameters before merger.  For J0737-3039A/B, the relatively short synchronization timescale ($\sim$8.4$\times10^{3}$ yrs) suggests that tidal interactions are actively working towards aligning the spins with the orbit, a process potentially observable through long-term precision timing with next-generation like LIGO-Virgo-KAGRA (LVK) collaboration \citep{2017PhRvL.119n1101A, 2023ApJ...946...59N}. Detecting subtle variation in spin or orbital parameters that arises from ongoing tidal synchronization offers  valuable constraints for probing the   internal physics neutron star and the mechanisms that govern  tidal dissipation.

The apsidal motion constant, $k_{2}$, serves as  a direct link to the  neutron star's equation of state (EoS) \citep{1994ApJ...424..823C}, and references therein. The value adopted in this study,  $k_2 = 0.1$ aligns with prediction from several theoretical EoS models (\citep{2021PhRvX..11d1050K, 2010PhRvD..81l3016H}). Nevertheless, precise measurements of apsidal motion, particularly  non-GR contributions, could help discriminate between different EoS predictions\citep{2002Sci...298.1592D, 2016PhR...621..127L}. Future observational improvements might allow for the disentangling of the oblateness and tidal contributions from the dominant GR precession, offering a powerful probe of the NS structure.

It is useful to recognize what the equilibrium tide model can and cannot tell us. The model we use (Zahn 1977, 1989) treats the star as if it reacts instantly to the tidal forces acting on it. In reality, especially in neutron stars, do not behave this simply. Their internal oscillation modes can be excited by the orbit's tidal forcing. These dynamics tides can increase energy dissipation  when the orbital frequency, or one of its harmonics, comes close to an oscillation mode such as an f-mode or g-mode (e.g., \citep{2021MNRAS.504.1273P, 2024PhRvD.110b4039Y}). This  can change the synchronization timescale, and in some case, leave a measurably imprint on the GW signal during the late inspiral phase (\citep{2021MNRAS.504.1273P, 2025LRR....28....3Y}). Future work should incorporate models of dynamic tides to provide a more complete picture of tidal evolution in these systems.

The study of tidal interactions and orbital evolution in binary pulsars has profound implications for multi-messenger astronomy. The tidal deformability of neutron stars, directly related to $k_{2}$ and the EoS, leaves an imprint on the GW form during the late inspiral phase of binary neutron star mergers, detectable by ground-based interferometers like LIGO, Virgo, and KAGRA (e.g., \citep{2016PhRvL.116x1103A, 2023PhRvX..13d1039A, 2023ApJ...946...59N}). Constraining tidal effects through pulsar timing complements GW observations, providing independent checks on the EoS in different regimes \citep{2002Sci...298.1592D, 2012ARNPS..62..485L}. Understanding the pre-merger evolution, including synchronization state and eccentricity, is crucial for accurately modeling merger dynamics and interpreting electromagnetic counterparts.



Finally, the periastron advance rates measured in these systems match the predictions of GR with remarkable precision. For PSR 1913+16, the observed rate of periastron advance is 4.2 degrees per year, in excellent agreement with the general relativistic prediction.


\section{Conclusions}
This study analyzed  the apsidal motion and tidal evolution of three non-synchronous binary pulsar systems: PSR 1913+16 (Hulse-Taylor pulsar), J0737-3039A/B (double pulsar), and J0621+1002 (intermediate-mass binary pulsar). Using Zahn’s tidal equations and integrating them with a classical fourth-order Runge-Kutta Scheme, we quantified the relative contributions of GR, stellar oblateness, and tidal interactions to the orbital and spin evolution of these binaries Our key findings can be summarized as follows:

1. The dominance of GW effects: The orbital evolution in PSR 1913+16 and J0737-3039A/B, is primarily driven by GW emission rather than tidal interactions. The observed orbital period decay of 76.5 $\mu$s/yr in PSR 1913+16 due to gravitational radiation far exceeds the changes predicted by tidal effects alone.


2. System-dependent evolution timescales: The synchronization and circularization timescales vary significantly among the three binary pulsar systems. In PSR B1913+16, synchronization occurs in 
$\simeq 1.2\times10^{5}$ yrs and circularization in $\simeq3.5\times 10^{7}$ yrs, both well before its 300 Myr merger time. For J0737–3039A/B, shows the most rapid evolution with synchronization achieved in approximately $\simeq8.4\times 10^{3}$ yrs and circularizes in $\simeq1.1\times 10^{6}$ yrs, much earlier than its 85 Myr merger. In contrast, PSR J0621+1002 shows minimal orbital and spin evolution due to tides or GW emission over astrophysically relevant timescales $\sim10^{10}$ yrs.


3. Apsidal motion as a relativistic probe: The periastron advance rates measured in these systems match the predictions of GR with remarkable precision, confirming Einstein's theory in strong-field regimes. The agreement between observed and predicted values for multiple post-Keplerian parameters in the double pulsar system J0737-3039A/B provides one of the most stringent tests of general relativity to date.

4. Evolutionary Patterns: Our evolutionary curves reveal distinct patterns in semi-major axis reduction, eccentricity decay, and spin acceleration across the three systems. These patterns reflect the underlying physical processes and system parameters, particularly the orbital separation, component masses, and companion types.

5. Tidal parameters: The derived apsidal motion constants ($k_{2}\approx0.1$) and tidal friction times ($t_{F}$ ranging from hours to days) align with theoretical predictions for neutron stars and provide valuable constraints on stellar structure models.

The study of apsidal motion in non-synchronous binary pulsars continues to yield valuable insights into relativistic gravity, stellar structure, and orbital dynamics. These systems serve as exceptional astrophysical laboratories for testing fundamental physics in extreme environments. Future observations with increased precision, particularly of the double pulsar system J0737-3039A/B, will further constrain theoretical models and potentially reveal subtle deviations from general relativity that could point toward new physics.

Future work should focus on integrating GW emission terms with tidal effects for a more comprehensive model of binary pulsar evolution. Additionally, expanding the analysis to include more binary pulsar systems with diverse orbital parameters would provide a broader dataset for testing evolutionary models and refining our understanding of these fascinating astrophysical systems. More sophisticated models incorporating additional physical effects could yield more accurate results.

 \section{Data availability} The research presented here has made extensive use of the 2025 version of the ATNF Pulsar Catalogue \citep{2005AJ....129.1993M}. All data used in this study are freely available at
https://www.atnf.csiro.au/research/pulsar/psrcat/. 
 \section{Declarations}
\textbf{Ethical Statement} There is no ethical issue, presently, the manuscript
is submitted in this journal. 
 \section{Competing interests} The author declares that there are no competing interests.

 \section{Funding}
No funding was received for conducting this study.

 \section{Acknowledgments}
The author thanks the anonymous referee
for the careful reading of the manuscript and for all suggestions and comments which
allowed us to improve both the quality and the clarity of the paper.

\backmatter




\bibliography{sn-bibliography}

@ARTICLE{2025MPLA...4050101A,
       author = {{Abu-Saleem}, Mohammed and {Taani}, Ali},
        title = "{Topological aspects of double neutron star formation: Its application to the gravitational-wave signatures}",
      journal = {Modern Physics Letters A},
         year = 2025,
        month = aug,
       volume = {40},
       number = {26},
          eid = {2550101},
        pages = {2550101},
          doi = {10.1142/S0217732325501019},
       adsurl = {https://ui.adsabs.harvard.edu/abs/2025MPLA...4050101A}
}

@ARTICLE{2024PhRvD.110b4039Y,
       author = {{Yu}, Hang and {Arras}, Phil and {Weinberg}, Nevin N.},
        title = "{Dynamical tides during the inspiral of rapidly spinning neutron stars: Solutions beyond mode resonance}",
      journal = {Physical Review D},
         year = 2024,
        month = jul,
       volume = {110},
       number = {2},
          eid = {024039},
        pages = {024039},
          doi = {10.1103/PhysRevD.110.024039},
archivePrefix = {arXiv},
       eprint = {2404.00147},
 primaryClass = {gr-qc},
       adsurl = {https://ui.adsabs.harvard.edu/abs/2024PhRvD.110b4039Y}
}

@BOOK{1972gcpa.book.....W,
       author = {{Weinberg}, Steven},
        title = "{Gravitation and Cosmology: Principles and Applications of the General Theory of Relativity}",
         year = 1972,
       adsurl = {https://ui.adsabs.harvard.edu/abs/1972gcpa.book.....W}
}

@ARTICLE{1989A&A...223..112Z,
       author = {{Zahn}, J. -P. and {Bouchet}, L.},
        title = "{Tidal evolution of close binary stars. II. Orbital circularization oflate-type binaries.}",
      journal = {aap},
         year = 1989,
        month = oct,
       volume = {223},
        pages = {112-118},
       adsurl = {https://ui.adsabs.harvard.edu/abs/1989A&A...223..112Z}
}

@INPROCEEDINGS{1976IAUS...73...75P,
       author = {{Paczynski}, B.},
        title = "{Common Envelope Binaries}",
    booktitle = {Structure and Evolution of Close Binary Systems},
         year = 1976,
       editor = {{Eggleton}, Peter and {Mitton}, Simon and {Whelan}, John},
       series = {IAU Symposium},
       volume = {73},
        month = jan,
        pages = {75},
       adsurl = {https://ui.adsabs.harvard.edu/abs/1976IAUS...73...75P}
}

@ARTICLE{2012NewAR..56..122W,
       author = {{Wang}, Bo and {Han}, Zhanwen},
        title = "{Progenitors of type Ia supernovae}",
      journal = {nar},
         year = 2012,
        month = jun,
       volume = {56},
       number = {4},
        pages = {122-141},
          doi = {10.1016/j.newar.2012.04.001},
archivePrefix = {arXiv},
       eprint = {1204.1155},
 primaryClass = {astro-ph.SR},
       adsurl = {https://ui.adsabs.harvard.edu/abs/2012NewAR..56..122W}
}

@BOOK{1973bmss.book.....B,
       author = {{Batten}, Alan H.},
        title = "{Binary and multiple systems of stars}",
         year = 1973,
publisher=Wiley,
       adsurl = {https://ui.adsabs.harvard.edu/abs/1973bmss.book.....B}
}

@ARTICLE{2006Sci...314...97K,
       author = {{Kramer}, M. and {Stairs}, I.~H. and {Manchester}, R.~N. and {McLaughlin}, M.~A. and {Lyne}, A.~G. and {Ferdman}, R.~D. and {Burgay}, M. and {Lorimer}, D.~R. and {Possenti}, A. and {D'Amico}, N. and {Sarkissian}, J.~M. and {Hobbs}, G.~B. and {Reynolds}, J.~E. and {Freire}, P.~C.~C. and {Camilo}, F.},
        title = "{Tests of General Relativity from Timing the Double Pulsar}",
      journal = {Science},
         year = 2006,
        month = oct,
       volume = {314},
       number = {5796},
        pages = {97-102},
          doi = {10.1126/science.1132305},
archivePrefix = {arXiv},
       eprint = {astro-ph/0609417},
 primaryClass = {astro-ph},
       adsurl = {https://ui.adsabs.harvard.edu/abs/2006Sci...314...97K}
}

@ARTICLE{2021PhRvX..11d1050K,
       author = {{Kramer}, M. and {Stairs}, I.~H. and {Manchester}, R.~N. and {Wex}, N. and {Deller}, A.~T. and {Coles}, W.~A. and {Ali}, M. and {Burgay}, M. and {Camilo}, F. and {Cognard}, I. and {Damour}, T. and {Desvignes}, G. and {Ferdman}, R.~D. and {Freire}, P.~C.~C. and {Grondin}, S. and {Guillemot}, L. and {Hobbs}, G.~B. and {Janssen}, G. and {Karuppusamy}, R. and {Lorimer}, D.~R. and {Lyne}, A.~G. and {McKee}, J.~W. and {McLaughlin}, M. and {M{\"u}nch}, L.~E. and {Perera}, B.~B.~P. and {Pol}, N. and {Possenti}, A. and {Sarkissian}, J. and {Stappers}, B.~W. and {Theureau}, G.},
        title = "{Strong-Field Gravity Tests with the Double Pulsar}",
      journal = {Physical Review X},
         year = 2021,
        month = oct,
       volume = {11},
       number = {4},
          eid = {041050},
        pages = {041050},
          doi = {10.1103/PhysRevX.11.041050},
archivePrefix = {arXiv},
       eprint = {2112.06795},
 primaryClass = {astro-ph.HE},
       adsurl = {https://ui.adsabs.harvard.edu/abs/2021PhRvX..11d1050K}
}

@ARTICLE{2021MNRAS.504.1273P,
       author = {{Passamonti}, A. and {Andersson}, N. and {Pnigouras}, P.},
        title = "{Dynamical tides in neutron stars: the impact of the crust}",
      journal = {MNRAS},
         year = 2021,
        month = jun,
       volume = {504},
       number = {1},
        pages = {1273-1293},
          doi = {10.1093/mnras/stab870},
archivePrefix = {arXiv},
       eprint = {2012.09637},
 primaryClass = {astro-ph.HE},
       adsurl = {https://ui.adsabs.harvard.edu/abs/2021MNRAS.504.1273P}
}

@INPROCEEDINGS{2007IAUS..240..290G,
       author = {{Gim{\'e}nez}, Alvaro},
        title = "{The Apsidal Motion Test in Eclipsing Binaries}",
    booktitle = {Binary Stars as Critical Tools \& Tests in Contemporary Astrophysics},
         year = 2007,
       editor = {{Hartkopf}, William I. and {Harmanec}, Petr and {Guinan}, Edward F.},
       series = {IAU Symposium},
       volume = {240},
        month = aug,
        pages = {290-298},
          doi = {10.1017/S1743921307004188},
       adsurl = {https://ui.adsabs.harvard.edu/abs/2007IAUS..240..290G}
}

@ARTICLE{2002MNRAS.330..435B,
       author = {{Benvenuto}, O.~G. and {Serenelli}, A.~M. and {Althaus}, L.~G. and {Barb{\'a}}, R.~H. and {Morrell}, N.~I.},
        title = "{Calculation of the masses of the binary star HD 93205 by application of the theory of apsidal motion}",
      journal = {mnras},
         year = 2002,
        month = feb,
       volume = {330},
       number = {2},
        pages = {435-442},
          doi = {10.1046/j.1365-8711.2002.05083.x},
archivePrefix = {arXiv},
       eprint = {astro-ph/0110662},
 primaryClass = {astro-ph},
       adsurl = {https://ui.adsabs.harvard.edu/abs/2002MNRAS.330..435B}
}

@ARTICLE{1939MNRAS..99..451S,
       author = {{Sterne}, T.~E.},
        title = "{Apsidal motion in binary stars}",
      journal = {mnras},
         year = 1939,
        month = mar,
       volume = {99},
        pages = {451-462},
          doi = {10.1093/mnras/99.5.451},
       adsurl = {https://ui.adsabs.harvard.edu/abs/1939MNRAS..99..451S}
}

@ARTICLE{2003LRR.....6....5S,
       author = {{Stairs}, Ingrid H.},
        title = "{Testing General Relativity with Pulsar Timing}",
      journal = {Living Reviews in Relativity},
         year = 2003,
        month = dec,
       volume = {6},
       number = {1},
          eid = {5},
        pages = {5},
          doi = {10.12942/lrr-2003-5},
archivePrefix = {arXiv},
       eprint = {astro-ph/0307536},
 primaryClass = {astro-ph},
       adsurl = {https://ui.adsabs.harvard.edu/abs/2003LRR.....6....5S}
}

@ARTICLE{2004Sci...304..547S,
       author = {{Stairs}, Ingrid H.},
        title = "{Pulsars in Binary Systems: Probing Binary Stellar Evolution and General Relativity}",
      journal = {Science},
         year = 2004,
        month = apr,
       volume = {304},
       number = {5670},
        pages = {547-552},
          doi = {10.1126/science.1096986},
       adsurl = {https://ui.adsabs.harvard.edu/abs/2004Sci...304..547S}
}

@ARTICLE{2025LRR....28....3Y,
       author = {{Yunes}, Nicol{\'a}s and {Siemens}, Xavier and {Yagi}, Kent},
        title = "{Gravitational-wave tests of general relativity with ground-based detectors and pulsar-timing arrays}",
      journal = {Living Reviews in Relativity},
         year = 2025,
        month = mar,
       volume = {28},
       number = {1},
          eid = {3},
        pages = {3},
          doi = {10.1007/s41114-024-00054-9},
archivePrefix = {arXiv},
       eprint = {2408.05240},
 primaryClass = {gr-qc},
       adsurl = {https://ui.adsabs.harvard.edu/abs/2025LRR....28....3Y}
}

@BOOK{1986bhwd.book.....S,
       author = {{Shapiro}, Stuart L. and {Teukolsky}, Saul A.},
        title = "{Black Holes, White Dwarfs and Neutron Stars: The Physics of Compact Objects}",
         year = 1986,
publisher=Wiley,
       adsurl = {https://ui.adsabs.harvard.edu/abs/1986bhwd.book.....S}
}

@ARTICLE{2021AN....342..625A,
       author = {{Almusleh}, Nour Aldein and {Taani}, Ali and {{O}zdemir}, Sergen and {Rah}, Maria and {Al-Wardat}, Mashhoor A. and {Zhao}, Gang and {Mardini}, Mohammad K.},
        title = "{Metal-poor stars observed with the automated planet finder telescope. III. CEMP-no stars are the descendant of population III stars}",
      journal = {Astronomische Nachrichten},
         year = 2021,
        month = may,
       volume = {342},
       number = {4},
        pages = {625-632},
          doi = {10.1002/asna.202113867},
archivePrefix = {arXiv},
       eprint = {2102.01076},
 primaryClass = {astro-ph.GA},
       adsurl = {https://ui.adsabs.harvard.edu/abs/2021AN....342..625A}
}

@ARTICLE{2009PhRvD..80h4035D,
       author = {{Damour}, Thibault and {Nagar}, Alessandro},
        title = "{Relativistic tidal properties of neutron stars}",
      journal = {prd},
         year = 2009,
        month = oct,
       volume = {80},
       number = {8},
          eid = {084035},
        pages = {084035},
          doi = {10.1103/PhysRevD.80.084035},
archivePrefix = {arXiv},
       eprint = {0906.0096},
 primaryClass = {gr-qc},
       adsurl = {https://ui.adsabs.harvard.edu/abs/2009PhRvD..80h4035D}
}

@ARTICLE{2017PhRvD..96d3002O,
       author = {{Ofengeim}, D.~D. and {Fortin}, M. and {Haensel}, P. and {Yakovlev}, D.~G. and {Zdunik}, J.~L.},
        title = "{Neutrino luminosities and heat capacities of neutron stars in analytic form}",
      journal = {prd},
         year = 2017,
        month = aug,
       volume = {96},
       number = {4},
          eid = {043002},
        pages = {043002},
          doi = {10.1103/PhysRevD.96.043002},
archivePrefix = {arXiv},
       eprint = {1708.08272},
 primaryClass = {astro-ph.HE},
       adsurl = {https://ui.adsabs.harvard.edu/abs/2017PhRvD..96d3002O}
}

@ARTICLE{2011A&A...527A..83Z,
       author = {{Zhang}, C.~M. and {Wang}, J. and {Zhao}, Y.~H. and {Yin}, H.~X. and {Song}, L.~M. and {Menezes}, D.~P. and {Wickramasinghe}, D.~T. and {Ferrario}, L. and {Chardonnet}, P.},
        title = "{Study of measured pulsar masses and their possible conclusions}",
      journal = {aap},
         year = 2011,
        month = mar,
       volume = {527},
          eid = {A83},
        pages = {A83},
          doi = {10.1051/0004-6361/201015532},
archivePrefix = {arXiv},
       eprint = {1010.5429},
 primaryClass = {astro-ph.HE},
       adsurl = {https://ui.adsabs.harvard.edu/abs/2011A&A...527A..83Z}
}

@ARTICLE{2025arXiv250502600N,
       author = {{Nava-Callejas}, Martin and {Page}, Dany and {Cavecchi}, Yuri},
        title = "{Thermonuclear Heating of Accreting Neutron Stars}",
      journal = {arXiv e-prints},
         year = 2025,
        month = may,
          eid = {arXiv:2505.02600},
        pages = {arXiv:2505.02600},
          doi = {10.48550/arXiv.2505.02600},
archivePrefix = {arXiv},
       eprint = {2505.02600},
 primaryClass = {astro-ph.HE},
       adsurl = {https://ui.adsabs.harvard.edu/abs/2025arXiv250502600N}
}

@ARTICLE{2010PhRvD..81l3016H,
       author = {{Hinderer}, Tanja and {Lackey}, Benjamin D. and {Lang}, Ryan N. and {Read}, Jocelyn S.},
        title = "{Tidal deformability of neutron stars with realistic equations of state and their gravitational wave signatures in binary inspiral}",
      journal = {prd},
         year = 2010,
        month = jun,
       volume = {81},
       number = {12},
          eid = {123016},
        pages = {123016},
          doi = {10.1103/PhysRevD.81.123016},
archivePrefix = {arXiv},
       eprint = {0911.3535},
 primaryClass = {astro-ph.HE},
       adsurl = {https://ui.adsabs.harvard.edu/abs/2010PhRvD..81l3016H}
}

@ARTICLE{2002A&A...388..518C,
       author = {{Claret}, A. and {Willems}, B.},
        title = "{New results on the apsidal-motion test to stellar structure and evolution including the effects of dynamic tides}",
      journal = {aap},
         year = 2002,
        month = jun,
       volume = {388},
        pages = {518-530},
          doi = {10.1051/0004-6361:20020425},
       adsurl = {https://ui.adsabs.harvard.edu/abs/2002A&A...388..518C}
}

@ARTICLE{2021arXiv210513335P,
       author = {{Poddar}, Tanmay Kumar and {Mohanty}, Subhendra and {Jana}, Soumya},
        title = "{Gravitational radiation from binary systems in massive graviton theories}",
      journal = {arXiv e-prints},
         year = 2021,
        month = may,
          eid = {arXiv:2105.13335},
        pages = {arXiv:2105.13335},
          doi = {10.48550/arXiv.2105.13335},
archivePrefix = {arXiv},
       eprint = {2105.13335},
 primaryClass = {gr-qc},
       adsurl = {https://ui.adsabs.harvard.edu/abs/2021arXiv210513335P}
}

@ARTICLE{2025PhLB..86439388B,
       author = {{Bonilla}, Alexander and {Santoni}, Alessandro and {Nunes}, Rafael C. and {Said}, Jackson Levi},
        title = "{VSL-Gravity in light of PSR B1913+16 full data set: Upper limits on graviton mass and its theoretical consequences}",
      journal = {Physics Letters B},
         year = 2025,
        month = may,
       volume = {864},
          eid = {139388},
        pages = {139388},
          doi = {10.1016/j.physletb.2025.139388},
archivePrefix = {arXiv},
       eprint = {2503.12195},
 primaryClass = {gr-qc},
       adsurl = {https://ui.adsabs.harvard.edu/abs/2025PhLB..86439388B}
}

@ARTICLE{2011PhRvL.107n1101L,
       author = {{Le Tiec}, Alexandre and {Mrou{\'e}}, Abdul H. and {Barack}, Leor and {Buonanno}, Alessandra and {Pfeiffer}, Harald P. and {Sago}, Norichika and {Taracchini}, Andrea},
        title = "{Periastron Advance in Black-Hole Binaries}",
      journal = {prl},
         year = 2011,
        month = sep,
       volume = {107},
       number = {14},
          eid = {141101},
        pages = {141101},
          doi = {10.1103/PhysRevLett.107.141101},
archivePrefix = {arXiv},
       eprint = {1106.3278},
 primaryClass = {gr-qc},
       adsurl = {https://ui.adsabs.harvard.edu/abs/2011PhRvL.107n1101L}
}

@ARTICLE{1999A&A...350..928T,
       author = {{Tauris}, Thomas M. and {Savonije}, Gerrit J.},
        title = "{Formation of millisecond pulsars. I. Evolution of low-mass X-ray binaries with P\_orb> 2 days}",
      journal = {aap},
         year = 1999,
        month = oct,
       volume = {350},
        pages = {928-944},
          doi = {10.48550/arXiv.astro-ph/9909147},
archivePrefix = {arXiv},
       eprint = {astro-ph/9909147},
 primaryClass = {astro-ph},
       adsurl = {https://ui.adsabs.harvard.edu/abs/1999A&A...350..928T}
}

@ARTICLE{2023A&A...672A...9M,
       author = {{Mor{\'a}n-Fraile}, Javier and {Schneider}, Fabian R.~N. and {R{\"o}pke}, Friedrich K. and {Ohlmann}, Sebastian T. and {Pakmor}, R{\"u}diger and {Soultanis}, Theodoros and {Bauswein}, Andreas},
        title = "{Gravitational wave emission from dynamical stellar interactions}",
      journal = {aap},
         year = 2023,
        month = apr,
       volume = {672},
          eid = {A9},
        pages = {A9},
          doi = {10.1051/0004-6361/202245109},
archivePrefix = {arXiv},
       eprint = {2303.05519},
 primaryClass = {astro-ph.HE},
       adsurl = {https://ui.adsabs.harvard.edu/abs/2023A&A...672A...9M}
}

@article{10.1093/mnras/sty2177,
    author = {Susobhanan, Abhimanyu and Gopakumar, Achamveedu and Joshi, Bhal Chandra and Kumar, Ranjan},
    title = {Exploring the effect of periastron advance in small-eccentricity binary pulsars},
    journal = {Monthly Notices of the Royal Astronomical Society},
    volume = {480},
    number = {4},
    pages = {5260-5271},
    year = {2018},
    month = {08},
    issn = {0035-8711},
    doi = {10.1093/mnras/sty2177},
    url = {https://doi.org/10.1093/mnras/sty2177},
    eprint = {https://academic.oup.com/mnras/article-pdf/480/4/5260/25636554/sty2177.pdf},
}

@ARTICLE{2022A&A...663A..75W,
       author = {{Wang}, D. and {Gong}, B.~P.},
        title = "{Possible tidal dissipation in millisecond pulsar binaries}",
      journal = {aap},
         year = 2022,
        month = jul,
       volume = {663},
          eid = {A75},
        pages = {A75},
          doi = {10.1051/0004-6361/201937415},
archivePrefix = {arXiv},
       eprint = {2207.08170},
 primaryClass = {astro-ph.HE},
       adsurl = {https://ui.adsabs.harvard.edu/abs/2022A&A...663A..75W}
}

@ARTICLE{2012MNRAS.421..426F,
       author = {{Fuller}, Jim and {Lai}, Dong},
        title = "{Dynamical tides in compact white dwarf binaries: tidal synchronization and dissipation}",
      journal = {mnras},
         year = 2012,
        month = mar,
       volume = {421},
       number = {1},
        pages = {426-445},
          doi = {10.1111/j.1365-2966.2011.20320.x},
archivePrefix = {arXiv},
       eprint = {1108.4910},
 primaryClass = {astro-ph.SR},
       adsurl = {https://ui.adsabs.harvard.edu/abs/2012MNRAS.421..426F}
}

@article{10.1093/mnras/stx540,
    author = {Lin, Yufeng and Ogilvie, Gordon I.},
    title = {Tidal interactions in spin–orbit misaligned systems},
    journal = {Monthly Notices of the Royal Astronomical Society},
    volume = {468},
    number = {2},
    pages = {1387-1397},
    year = {2017},
    month = {03},
    issn = {0035-8711},
    doi = {10.1093/mnras/stx540},
    url = {https://doi.org/10.1093/mnras/stx540},
    eprint = {https://academic.oup.com/mnras/article-pdf/468/2/1387/11126967/stx540.pdf},
}

@INPROCEEDINGS{2008EAS....29...67Z,
       author = {{Zahn}, J. -P.},
        title = "{Tidal dissipation in binary systems}",
    booktitle = {EAS Publications Series},
         year = 2008,
       editor = {{Goupil}, M. -J. and {Zahn}, J. -P.},
       series = {EAS Publications Series},
       volume = {29},
        month = jan,
        pages = {67-90},
          doi = {10.1051/eas:0829002},
archivePrefix = {arXiv},
       eprint = {0807.4870},
 primaryClass = {astro-ph},
       adsurl = {https://ui.adsabs.harvard.edu/abs/2008EAS....29...67Z}
}

@ARTICLE{2016ApJ...829...55W,
       author = {{Weisberg}, J.~M. and {Huang}, Y.},
        title = "{Relativistic Measurements from Timing the Binary Pulsar PSR B1913+16}",
      journal = {apj},
         year = 2016,

       volume = {829},
       number = {1},
          eid = {55},
        pages = {55},
         
archivePrefix = {arXiv},
       eprint = {1606.02744},
 primaryClass = {astro-ph.HE},
       adsurl = {https://ui.adsabs.harvard.edu/abs/2016ApJ...829...55W}
}

@ARTICLE{2010ApJ...722.1030W,
       author = {{Weisberg}, J.~M. and {Nice}, D.~J. and {Taylor}, J.~H.},
        title = "{Timing Measurements of the Relativistic Binary Pulsar PSR B1913+16}",
      journal = {apj},
         year = 2010,
    
       volume = {722},
       number = {2},
        pages = {1030-1034},
    
archivePrefix = {arXiv},
       eprint = {1011.0718},
 primaryClass = {astro-ph.GA},
       adsurl = {https://ui.adsabs.harvard.edu/abs/2010ApJ...722.1030W}
}

@ARTICLE{1980ApL....21...79K,
       author = {{Kopal}, Z. and {Kondo}, Y.},
        title = "{Book-Review - Dynamics of Close Binary Systems}",
      journal = {Applied Letters},
         year = 1980,
        month = jan,
       volume = {21},
        pages = {79},
       adsurl = {https://ui.adsabs.harvard.edu/abs/1980ApL....21...79K}
}

@ARTICLE{1996ApJ...461..812C,
       author = {{Camilo}, F. and {Nice}, D.~J. and {Taylor}, J.~H.},
        title = "{A Search for Millisecond Pulsars at Galactic Latitudes -50 degrees < B < -20 degrees}",
      journal = {apj},
         year = 1996,
        month = apr,
       volume = {461},
        pages = {812},
          doi = {10.1086/177103},
       adsurl = {https://ui.adsabs.harvard.edu/abs/1996ApJ...461..812C}
}

@ARTICLE{2018JKAS...51....1L,
       author = {{Li}, Lin-Sen},
        title = "{Evolution of Orbit and Rotation of a Pseudo-Synchronous Binary System on the Main Sequence}",
      journal = {Journal of Korean Astronomical Society},
         year = 2018,
        month = feb,
       volume = {51},
       number = {1},
        pages = {1-4},
          doi = {10.5303/JKAS.2018.51.1.1},
       adsurl = {https://ui.adsabs.harvard.edu/abs/2018JKAS...51....1L}
}

@ARTICLE{2003Natur.426..531B,
       author = {{Burgay}, M. and {D'Amico}, N. and {Possenti}, A. and {Manchester}, R.~N. and {Lyne}, A.~G. and {Joshi}, B.~C. and {McLaughlin}, M.~A. and {Kramer}, M. and {Sarkissian}, J.~M. and {Camilo}, F. and {Kalogera}, V. and {Kim}, C. and {Lorimer}, D.~R.},
        title = "{An increased estimate of the merger rate of double neutron stars from observations of a highly relativistic system}",
      journal = {nat},
         year = 2003,
        month = dec,
       volume = {426},
       number = {6966},
        pages = {531-533},
          doi = {10.1038/nature02124},
archivePrefix = {arXiv},
       eprint = {astro-ph/0312071},
 primaryClass = {astro-ph},
       adsurl = {https://ui.adsabs.harvard.edu/abs/2003Natur.426..531B}
}

@ARTICLE{1975ApJ...195L..51H,
       author = {{Hulse}, R.~A. and {Taylor}, J.~H.},
        title = "{Discovery of a pulsar in a binary system.}",
      journal = {apjl},
         year = 1975,
        month = jan,
       volume = {195},
        pages = {L51-L53},
          doi = {10.1086/181708},
       adsurl = {https://ui.adsabs.harvard.edu/abs/1975ApJ...195L..51H}
}

@ARTICLE{2019A&A...628A..29C,
       author = {{Claret}, A.},
        title = "{Updating the theoretical tidal evolution constants: Apsidal motion and the moment of inertia}",
      journal = {aap},
         year = 2019,
        month = aug,
       volume = {628},
          eid = {A29},
        pages = {A29},
          doi = {10.1051/0004-6361/201936007},
archivePrefix = {arXiv},
       eprint = {1907.11538},
 primaryClass = {astro-ph.SR},
       adsurl = {https://ui.adsabs.harvard.edu/abs/2019A&A...628A..29C}
}

@ARTICLE{2025RNAAS...9...12W,
       author = {{Woodruff}, Hannah C. and {Samaraweera}, Dumindu and {Aufdenberg}, Jason P.},
        title = "{Exploring Spica's Apsidal Motion: Insights from Feature Engineering and Machine Learning with MESA}",
      journal = {Research Notes of the American Astronomical Society},
         year = 2025,
        month = jan,
       volume = {9},
       number = {1},
          eid = {12},
        pages = {12},
          doi = {10.3847/2515-5172/ada94c},
       adsurl = {https://ui.adsabs.harvard.edu/abs/2025RNAAS...9...12W}
}

@ARTICLE{1977A&A....57..383Z,
       author = {{Zahn}, J. -P.},
        title = "{Tidal friction in close binary systems.}",
      journal = {aap},
         year = 1977,
        month = may,
       volume = {57},
        pages = {383-394},
       adsurl = {https://ui.adsabs.harvard.edu/abs/1977A&A....57..383Z}
}

@ARTICLE{2010A&A...519A..57C,
       author = {{Claret}, A. and {Gim{\'e}nez}, A.},
        title = "{The apsidal-motion test of stellar structure and evolution: an update}",
      journal = {aap},
         year = 2010,
        month = sep,
       volume = {519},
          eid = {A57},
        pages = {A57},
          doi = {10.1051/0004-6361/201014008},
       adsurl = {https://ui.adsabs.harvard.edu/abs/2010A&A...519A..57C}
}

@ARTICLE{1993A&A...277..487C,
       author = {{Claret}, A. and {Gimenez}, A.},
        title = "{The apsidal motion test of the internal stellar structure:comparison between theory and observations.}",
      journal = {aap},
         year = 1993,
        month = oct,
       volume = {277},
        pages = {487-502},
       adsurl = {https://ui.adsabs.harvard.edu/abs/1993A&A...277..487C}
}

@ARTICLE{1989A&A...220..112Z,
       author = {{Zahn}, J. -P.},
        title = "{Tidal evolution of close binary stars. I - Revisiting the theory of the equilibrium tide}",
      journal = {aap},
         year = 1989,
        month = aug,
       volume = {220},
       number = {1-2},
        pages = {112-116},
       adsurl = {https://ui.adsabs.harvard.edu/abs/1989A&A...220..112Z}
}

@ARTICLE{2023A&A...678A.187C,
              author = {{Colom Bernadich}, M. and {Balakrishnan}, V. and {Barr}, E. and others},
        title = "{The MPIfR-MeerKAT Galactic Plane Survey. II. The eccentric double neutron star system PSR J1208{\ensuremath{-}}5936 and a neutron star merger rate update}",
      journal = {aap},
         year = 2023,
   
       volume = {678},
          eid = {A187},
        pages = {A187},
         
archivePrefix = {arXiv},
       eprint = {2308.16802},
 primaryClass = {astro-ph.HE},
       adsurl = {https://ui.adsabs.harvard.edu/abs/2023A&A...678A.187C}
}

@ARTICLE{2025Ap&SS.370...42T,
       author = {{Taani}, Ali and {Abu-Saleem}, Mohammed and {Mardini}, Mohammad and {Aljboor}, Hussam and {Tayem}, Mohammad},
        title = "{Exploring the formation mechanisms of double neutron star systems: an analytical perspective}",
      journal = {apss},
         year = 2025,
        month = may,
       volume = {370},
       number = {5},
          eid = {42},
        pages = {42},
          doi = {10.1007/s10509-025-04433-8},
archivePrefix = {arXiv},
       eprint = {2505.04778},
 primaryClass = {astro-ph.HE},
       adsurl = {https://ui.adsabs.harvard.edu/abs/2025Ap&SS.370...42T}
}

@ARTICLE{2020ApJ...892L...3A,
       author = {{Abbott}, B.~P. and {Abbott}, R. and {Abbott}, T.~D. and et al.},
        title = "{GW190425: Observation of a Compact Binary Coalescence with Total Mass {\ensuremath{\sim}} 3.4 M$_{{\ensuremath{\odot}}}$}",
      journal = {apjl},
         year = 2020,
  
       volume = {892},
       number = {1},
          eid = {L3},
        pages = {L3},
     
archivePrefix = {arXiv},
       eprint = {2001.01761},
 primaryClass = {astro-ph.HE}
}

@ARTICLE{2009ApJ...698..715S,
       author = {{Sirotkin}, Fedir V. and {Kim}, Woong-Tae},
        title = "{Internal Structure and Apsidal Motions of Polytropic Stars in Close Binaries}",
      journal = {apj},
         year = 2009,
        month = jun,
       volume = {698},
       number = {1},
        pages = {715-734},
          doi = {10.1088/0004-637X/698/1/715},
archivePrefix = {arXiv},
       eprint = {0904.2939},
 primaryClass = {astro-ph.SR},
       adsurl = {https://ui.adsabs.harvard.edu/abs/2009ApJ...698..715S}
}

@ARTICLE{1988ApJ...324L..71T,
       author = {{Tassoul}, Jean-Louis},
        title = "{On Orbital Circularization in Detached Close Binaries}",
      journal = {apjl},
         year = 1988,
        month = jan,
       volume = {324},
        pages = {L71},
          doi = {10.1086/185094},
       adsurl = {https://ui.adsabs.harvard.edu/abs/1988ApJ...324L..71T}
}

@ARTICLE{2024A&A...687A.167C,
       author = {{Claret}, A.},
        title = "{An approach to the effects of stellar rotation on the theoretical apsidal motion constants. Calculations from 0.40 M$_{{\ensuremath{\odot}}}$ to 25.0 M$_{{\ensuremath{\odot}}}$}",
      journal = {aap},
         year = 2024,
        month = jul,
       volume = {687},
          eid = {A167},
        pages = {A167},
          doi = {10.1051/0004-6361/202449379},
       adsurl = {https://ui.adsabs.harvard.edu/abs/2024A&A...687A.167C}
}

@ARTICLE{2024MNRAS.530.2822D,
       author = {{Dong}, Wenhao and {Melatos}, Andrew},
        title = "{Gravitational waves from non-radial oscillations of stochastically accreting neutron stars}",
      journal = {mnras},
         year = 2024,
        month = may,
       volume = {530},
       number = {3},
        pages = {2822-2839},
          doi = {10.1093/mnras/stae1028},
archivePrefix = {arXiv},
       eprint = {2404.11866},
 primaryClass = {astro-ph.HE},
       adsurl = {https://ui.adsabs.harvard.edu/abs/2024MNRAS.530.2822D}
}

@ARTICLE{2007MNRAS.382..356K,
       author = {{Khaliullin}, Kh. F. and {Khaliullina}, A.~I.},
        title = "{Determination of the axial rotation rate using apsidal motion for early-type eclipsing binaries}",
      journal = {mnras},
         year = 2007,
        month = nov,
       volume = {382},
       number = {1},
        pages = {356-366},
          doi = {10.1111/j.1365-2966.2007.12375.x},
       adsurl = {https://ui.adsabs.harvard.edu/abs/2007MNRAS.382..356K}
}

@ARTICLE{2022PhRvD.106f3005K,
       author = {{Kunjipurayil}, Athul and {Zhao}, Tianqi and {Kumar}, Bharat and {Agrawal}, Bijay K. and {Prakash}, Madappa},
        title = "{Impact of the equation of state on f - and p - mode oscillations of neutron stars}",
      journal = {prd},
         year = 2022,
        month = sep,
       volume = {106},
       number = {6},
          eid = {063005},
        pages = {063005},
          doi = {10.1103/PhysRevD.106.063005},
archivePrefix = {arXiv},
       eprint = {2205.02081},
 primaryClass = {nucl-th},
       adsurl = {https://ui.adsabs.harvard.edu/abs/2022PhRvD.106f3005K}
}

@ARTICLE{2016PhRvL.116x1103A,
       author = {{Abbott}, B.~P. and {Abbott}, R. and {Abbott}, T.~D. and et al.},
        title = "{GW151226: Observation of Gravitational Waves from a 22-Solar-Mass Binary Black Hole Coalescence}",
      journal = {prl},
         year = 2016,

       volume = {116},
       number = {24},
          eid = {241103},
        pages = {241103},
      
archivePrefix = {arXiv},
       eprint = {1606.04855},
 primaryClass = {gr-qc},
       adsurl = {https://ui.adsabs.harvard.edu/abs/2016PhRvL.116x1103A}
}

@ARTICLE{2017PhRvL.119n1101A,
       author = {{Abbott}, B.~P. and {Abbott}, R. and {Abbott}, T.~D and et al.},
        title = "{GW170814: A Three-Detector Observation of Gravitational Waves from a Binary Black Hole Coalescence}",
      journal = {prl},
         year = 2017,
 
       volume = {119},
       number = {14},
          eid = {141101},
        pages = {141101},
        
archivePrefix = {arXiv},
       eprint = {1709.09660},
 primaryClass = {gr-qc},
       adsurl = {https://ui.adsabs.harvard.edu/abs/2017PhRvL.119n1101A}
}

@ARTICLE{2012ChA&A..36..137C,
       author = {{Cai}, Yan and {Ali}, Taani and {Zhao}, Yong-heng and {Zhang}, Cheng-min},
        title = "{Statistics and Evolution of Pulsars' Parameters}",
      journal = {caa},
         year = 2012,
        month = apr,
       volume = {36},
       number = {2},
        pages = {137-147},
          doi = {10.1016/j.chinastron.2012.04.003},
       adsurl = {https://ui.adsabs.harvard.edu/abs/2012ChA&A..36..137C}
}

@ARTICLE{2010Ap&SS.327...59L,
       author = {{Li}, Lin-Sen},
        title = "{Post-Newtonian effect on the variation of time of periastron passage of binary stars in three gravitational theories}",
      journal = {apss},
         year = 2010,
        month = may,
       volume = {327},
       number = {1},
        pages = {59-65},
          doi = {10.1007/s10509-010-0267-4},
       adsurl = {https://ui.adsabs.harvard.edu/abs/2010Ap&SS.327...59L}
}

@ARTICLE{2011Ap&SS.334..125L,
       author = {{Li}, Lin-Sen},
        title = "{Influence of the gravitational radiation damping on the time of periastron passage of binary stars}",
      journal = {apss},
         year = 2011,
        month = jul,
       volume = {334},
       number = {1},
        pages = {125-130},
          doi = {10.1007/s10509-011-0693-y},
       adsurl = {https://ui.adsabs.harvard.edu/abs/2011Ap&SS.334..125L}
}

@ARTICLE{2013ApJ...767...85F,
       author = {{Ferdman}, R.~D. and {Stairs}, I.~H. and {Kramer}, M and et al.},
        title = "{The Double Pulsar: Evidence for Neutron Star Formation without an Iron Core-collapse Supernova}",
      journal = {apj},
         year = 2013,
        
       volume = {767},
       number = {1},
          eid = {85},
        pages = {85},
     
archivePrefix = {arXiv},
       eprint = {1302.2914},
 primaryClass = {astro-ph.SR},
       adsurl = {https://ui.adsabs.harvard.edu/abs/2013ApJ...767...85F}
}

@ARTICLE{2023PhRvX..13d1039A,
       author = {{Abbott}, R. and et al.},
        title = "{GWTC-3: Compact Binary Coalescences Observed by LIGO and Virgo during the Second Part of the Third Observing Run}",
      journal = {Physical Review X},
         year = 2023,
        
       volume = {13},
       number = {4},
          eid = {041039},
        pages = {041039},
       
archivePrefix = {arXiv},
       eprint = {2111.03606},
 primaryClass = {gr-qc},
       adsurl = {https://ui.adsabs.harvard.edu/abs/2023PhRvX..13d1039A}
}

@ARTICLE{1994ApJ...424..823C,
       author = {{Cook}, Gregory B. and {Shapiro}, Stuart L. and {Teukolsky}, Saul A.},
        title = "{Rapidly Rotating Neutron Stars in General Relativity: Realistic Equations of State}",
      journal = {apj},
         year = 1994,
        month = apr,
       volume = {424},
        pages = {823},
          doi = {10.1086/173934},
       adsurl = {https://ui.adsabs.harvard.edu/abs/1994ApJ...424..823C}
}

@ARTICLE{2023ApJ...946...59N,
       author = {{Nitz}, Alexander H. and {Kumar}, Sumit and {Wang}, Yi-Fan and {Kastha}, Shilpa and {Wu}, Shichao and {Sch{\"a}fer}, Marlin and {Dhurkunde}, Rahul and {Capano}, Collin D.},
        title = "{4-OGC: Catalog of Gravitational Waves from Compact Binary Mergers}",
      journal = {apj},
         year = 2023,
   
       volume = {946},
       number = {2},
          eid = {59},
        pages = {59},
       
archivePrefix = {arXiv},
       eprint = {2112.06878},
 primaryClass = {astro-ph.HE},
       adsurl = {https://ui.adsabs.harvard.edu/abs/2023ApJ...946...59N}
}

@ARTICLE{2005AJ....129.1993M,
       author = {{Manchester}, R.~N. and {Hobbs}, G.~B. and {Teoh}, A. and {Hobbs}, M.},
        title = "{The Australia Telescope National Facility Pulsar Catalogue}",
      journal = {aj},
         year = 2005,
 
       volume = {129},
       number = {4},
        pages = {1993-2006},
       
archivePrefix = {arXiv},
       eprint = {astro-ph/0412641},
 primaryClass = {astro-ph},
       adsurl = {https://ui.adsabs.harvard.edu/abs/2005AJ....129.1993M}
}

@ARTICLE{2017ChPhL..34l9701Y,
       author = {{Yang}, Yi-Yan and {Chen}, Li and {Linghu}, Rong-Feng and
         {Zhang}, Li-Yun and {Taani}, Ali},
        title = "{Constraints on Estimation of Radius of Double Pulsar PSR J0737-3039A and Its Neutron Star Nuclear Matter Composition}",
      journal = {Chinese Physics Letters},
         year = 2017,
  
       volume = {34},
       number = {12},
          eid = {129701},
        pages = {129701},
          
       adsurl = {https://ui.adsabs.harvard.edu/abs/2017ChPhL..34l9701Y}
}

@ARTICLE{2002Sci...298.1592D,
       author = {{Danielewicz}, Pawe{\l} and {Lacey}, Roy and {Lynch}, William G.},
        title = "{Determination of the Equation of State of Dense Matter}",
      journal = {Science},
         year = 2002,
        month = nov,
       volume = {298},
       number = {5598},
        pages = {1592-1596},
          doi = {10.1126/science.1078070},
archivePrefix = {arXiv},
       eprint = {nucl-th/0208016},
 primaryClass = {nucl-th},
       adsurl = {https://ui.adsabs.harvard.edu/abs/2002Sci...298.1592D}
}

@ARTICLE{2012ARNPS..62..485L,
       author = {{Lattimer}, James M.},
        title = "{The Nuclear Equation of State and Neutron Star Masses}",
      journal = {Annual Review of Nuclear and Particle Science},
         year = 2012,
        month = nov,
       volume = {62},
       number = {1},
        pages = {485-515},
          doi = {10.1146/annurev-nucl-102711-095018},
archivePrefix = {arXiv},
       eprint = {1305.3510},
 primaryClass = {nucl-th},
       adsurl = {https://ui.adsabs.harvard.edu/abs/2012ARNPS..62..485L}
}

@ARTICLE{2016PhR...621..127L,
       author = {{Lattimer}, James M. and {Prakash}, Madappa},
        title = "{The equation of state of hot, dense matter and neutron stars}",
      journal = {physrep},
         year = 2016,
        month = mar,
       volume = {621},
        pages = {127-164},
          doi = {10.1016/j.physrep.2015.12.005},
archivePrefix = {arXiv},
       eprint = {1512.07820},
 primaryClass = {astro-ph.SR},
       adsurl = {https://ui.adsabs.harvard.edu/abs/2016PhR...621..127L}
}

@ARTICLE{2023ChPhC..47d1002T,
       author = {{Taani}, Ali},
        title = "{Binding energy produced within the framework of the accretion of millisecond pulsars}",
      journal = {Chinese Physics C},
         year = 2023,
     
       volume = {47},
       number = {4},
          eid = {041002},
        pages = {041002},
    
archivePrefix = {arXiv},
       eprint = {2301.05928},
 primaryClass = {astro-ph.HE},
       adsurl = {https://ui.adsabs.harvard.edu/abs/2023ChPhC..47d1002T}
}

@ARTICLE{2023Galax..11...44T,
       author = {{Taani}, Ali},
        title = "{Characterizing the Regular Orbits of Binary Pulsars: An Initial Prospection Study}",
      journal = {Galaxies},
         year = 2023,

       volume = {11},
       number = {2},
        pages = {44},
        
       adsurl = {https://ui.adsabs.harvard.edu/abs/2023Galax..11...44T}
}

@ARTICLE{2022PASA...39...40T,
       author = {{Taani}, Ali and {Karino}, Shigeyuki and {Song}, Liming and {Zhang}, Chengmin and {Chaty}, Sylvain},
        title = "{Determination of wind-fed model parameters of neutron stars in high-mass X-ray binaries}",
      journal = {pasa},
         year = 2022,
     
       volume = {39},
          eid = {e040},
        pages = {e040},
         
       adsurl = {https://ui.adsabs.harvard.edu/abs/2022PASA...39...40T}
}

@ARTICLE{2022JHEAp..35...83T,
       author = {{Taani}, Ali and {Vallejo}, Juan C. and {Abu-Saleem}, Mohammed},
        title = "{Assessing the complexity of orbital parameters after asymmetric kick in binary pulsars}",
      journal = {Journal of High Energy Astrophysics},
         year = 2022,

       volume = {35},
        pages = {83-90},
         
archivePrefix = {arXiv},
       eprint = {2206.11015},
 primaryClass = {astro-ph.HE},
       adsurl = {https://ui.adsabs.harvard.edu/abs/2022JHEAp..35...83T}
}

@ARTICLE{2021AIPA...11a5309A,
       author = {{Abu-Saleem}, M. and {Taani}, A.},
        title = "{Retraction and folding on the hyperbolic black hole}",
      journal = {AIP Advances},
         year = 2021,
  
       volume = {11},
       number = {1},
          eid = {015309},
        pages = {015309},
        
       adsurl = {https://ui.adsabs.harvard.edu/abs/2021AIPA...11a5309A}
}

@ARTICLE{2017arXiv170204419T,
       author = {{Taani}, Ali},
        title = "{On the Distribution of Massive White Dwarfs and its Implication for Accretion-Induced Collapse}",
      journal = {arXiv e-prints},
         year = 2017,
        month = feb,
          eid = {arXiv:1702.04419},
        pages = {arXiv:1702.04419},
          doi = {10.48550/arXiv.1702.04419},
archivePrefix = {arXiv},
       eprint = {1702.04419},
 primaryClass = {astro-ph.HE},
       adsurl = {https://ui.adsabs.harvard.edu/abs/2017arXiv170204419T}
}

@ARTICLE{2017PASA...34...24T,
       author = {{Taani}, Ali and {Vallejo}, Juan C.},
        title = "{Dynamical Monte Carlo Simulations of 3-D Galactic Systems in Axisymmetric and Triaxial Potentials}",
      journal = {pasa},
         year = 2017,
     
       volume = {34},
          eid = {e024},
        pages = {e024},
     
archivePrefix = {arXiv},
       eprint = {1704.06412},
 primaryClass = {astro-ph.HE},
       adsurl = {https://ui.adsabs.harvard.edu/abs/2017PASA...34...24T}
}

@ARTICLE{2023RAA....23g5018A,
       author = {{Aljboor}, Hussam and {Taani}, Ali},
        title = "{Speckle-interferometric Study of Close Visual Binary System HIP 11253 (HD 14874) using Gaia (DR2 and EDR3)}",
      journal = {Research in Astronomy and Astrophysics},
         year = 2023,

       volume = {23},
       number = {7},
          eid = {075018},
        pages = {075018},
       
archivePrefix = {arXiv},
       eprint = {2301.05968},
 primaryClass = {astro-ph.SR},
       adsurl = {https://ui.adsabs.harvard.edu/abs/2023RAA....23g5018A}
}

@ARTICLE{2024NewA..10702149S,
       author = {{Saleem}, Mohammed Abu and {Taani}, Ali},
        title = "{The chaotic behavior of black holes: Investigating a topological retraction in anti-de Sitter spaces}",
      journal = {na},
         year = 2024,
        month = apr,
       volume = {107},
          eid = {102149},
        pages = {102149},
          doi = {10.1016/j.newast.2023.102149},
       adsurl = {https://ui.adsabs.harvard.edu/abs/2024NewA..10702149S}
}

\end{document}